\documentclass{cargese}\usepackage{graphicx}
\usepackage{amssymb}

\let\footnote\savefootnote
\let\footnotetext\savefootnotetext 
\let\footnoterule\savefootnoterule 
\normallatexbib

\begin{document}

\articletitle[Defects, Decay,\\ and Dissipated States]{
Defects, Decay, and Dissipated States}


\author{Emil J. Martinec} 



\affil{Enrico Fermi Institute and Department of Physics\\
        University of Chicago\\
        5640 S. Ellis Ave.\\
        Chicago, IL 60637-1433 }

\email{e-martinec@uchicago.edu}






\overfullrule=0pt
%
%
\def\nextline{\hfil\break}

\newcommand{\be}{\begin{equation}}
\newcommand{\ee}{\end{equation}}
\newcommand{\bbb}{\begin{eqnarray}}
\newcommand{\eee}{\end{eqnarray}}
\newcommand{\pref}[1]{(\ref{#1})}

\font\manual=manfnt \def\dbend{\lower3.5pt\hbox{\manual\char127}}
\def\danger#1{\smallskip\noindent\rlap\dbend%
\indent{\bf #1}\par\vskip-1.5pt\nobreak}

\def\ie{{\it i.e.}}
\def\eg{{\it e.g.}}
\def\cf{{\it c.f.}}
\def\etal{{\it et.al.}}
\def\etc{{\it etc}}

\def\st{\scriptstyle}
\def\sst{\scriptscriptstyle}
\def\tst{{\textstyle}}
\def\frac#1#2{{#1\over#2}}
\def\coeff#1#2{{\textstyle{#1\over #2}}}
\def\half{\frac12}
\def\hf{{\textstyle\half}}
\def\d{\partial}
\def\p{\partial}

\def\inbar{\,\vrule height1.5ex width.4pt depth0pt}
\def\IR{{\mathbb R}}
\def\IC{{\mathbb C}}
\def\IQ{{\mathbb Q}}
\def\IH{{\mathbb H}}
\def\IN{{\mathbb N}}
\def\IP{{\mathbb P}}
\def\IZ{{\mathbb Z}}
\def\Z{{\IZ}}
\def\One{{1\hskip -3pt {\rm l}}}
\def\nth{n^{\rm th}}
\catcode`\@=11
\def\slash#1{\mathord{\mathpalette\c@ncel{#1}}}
\def\underrel#1\over#2{\mathrel{\mathop{\kern\z@#1}\limits_{#2}}}
\def\lapprox{{\underrel{\scriptstyle<}\over\sim}}
\def\lessapprox{{\buildrel{<}\over{\scriptstyle\sim}}}
\catcode`\@=12
%
\def\sdtimes{\mathbin{\hbox{ \hskip-3pt $\times$ \hskip-5.3pt
\vrule height 4.5pt depth -.3pt width .35pt \hskip 2pt
}}}
%
%
\def\ket#1{|#1\rangle}
\def\bra#1{\langle#1|}
\def\vev#1{\langle#1\rangle}
\def\det{{\rm det}}
\def\tr{{\rm tr}}
\def\Tr{{\rm Tr}}
\def\mod{{\rm mod}}
\def\sinh{{\rm sinh}} 	\def\sh{{\rm sh}}
\def\cosh{{\rm cosh}} 	\def\ch{{\rm ch}}
\def\tanh{{\rm tanh}}
\def\sgn{{\rm sgn}}
\def\det{{\rm det}}
\def\Det{{\rm Det}}
\def\exp{{\rm exp}}
\def\ker{{\mathop{\rm ker}}}
\def\coker{{\mathop {\rm coker}}}
\def\dim{{\mathop{\rm dim}}}
\def\codim{{\mathop{\rm codim}}}
\def\Aut{{\rm Aut}}
\def\Hom{{\rm Hom}}
%
%
\def\AA{{\cal A}} \def\CA{{\cal A}}
\def\BB{{\cal B}} \def\CB{{\cal B}}
\def\CC{{\cal C}} 
\def\DD{{\cal D}} \def\CD{{\cal D}}
\def\EE{{\cal E}} \def\CE{{\cal E}}
\def\FF{{\cal F}} \def\CF{{\cal F}}
\def\GG{{\cal G}} \def\CG{{\cal G}}
\def\HH{{\cal H}} \def\CH{{\cal H}}
\def\II{{\cal I}} \def\CI{{\cal I}}
\def\JJ{{\cal J}} \def\CJ{{\cal J}}
\def\KK{{\cal K}} \def\CK{{\cal K}}
\def\LL{{\cal L}} \def\CL{{\cal L}}
\def\MM{{\cal M}} \def\CM{{\cal M}}
\def\NN{{\cal N}} \def\CN{{\cal N}}
\def\OO{{\cal O}} \def\CO{{\cal O}}
\def\PP{{\cal P}} \def\CP{{\cal P}}
\def\QQ{{\cal Q}} \def\CQ{{\cal Q}}
\def\RR{{\cal R}} \def\CR{{\cal R}}
\def\SS{{\cal S}} \def\CS{{\cal S}}
\def\TT{{\cal T}} \def\CT{{\cal T}}
\def\UU{{\cal U}} \def\CU{{\cal U}}
\def\VV{{\cal V}} \def\CV{{\cal V}}
\def\WW{{\cal W}} \def\CW{{\cal W}}
\def\XX{{\cal X}} \def\CX{{\cal X}}
\def\YY{{\cal Y}} \def\CY{{\cal Y}}
\def\ZZ{{\cal Z}} \def\CZ{{\cal Z}}
\def\lam{\lambda}
\def\eps{\epsilon}
\def\vareps{\varepsilon}
%
%
 
\def\makeblankbox#1#2{\hbox{\lower\dp0\vbox{\hidehrule{#1}{#2}%
   \kern -#1
   \hbox to \wd0{\hidevrule{#1}{#2}%
      \raise\ht0\vbox to #1{}
      \lower\dp0\vtop to #1{}
      \hfil\hidevrule{#2}{#1}}%
   \kern-#1\hidehrule{#2}{#1}}}%
}%
\def\hidehrule#1#2{\kern-#1\hrule height#1 depth#2 \kern-#2}%
\def\hidevrule#1#2{\kern-#1{\dimen0=#1\advance\dimen0 by #2\vrule
    width\dimen0}\kern-#2}%
\def\openbox{\ht0=1.2mm \dp0=1.2mm \wd0=2.4mm  \raise 2.75pt
\makeblankbox {.25pt} {.25pt}  }
\def\qed{\hskip 8mm \openbox}
\def\abs#1{\left\vert #1 \right\vert}
\def\bun#1/#2{\leavevmode
   \kern.1em \raise .5ex \hbox{\the\scriptfont0 #1}%
   \kern-.1em $/$%
   \kern-.15em \lower .25ex \hbox{\the\scriptfont0 #2}%
}
\def\row#1#2{#1_1,\ldots,#1_#2}
\def\apar{\noalign{\vskip 2mm}}
\def\blackbox{\hbox{\vrule height .5ex width .3ex depth -.3ex}}
\def\nord{{\textstyle {\blackbox\atop\blackbox}}}
\def\ts{\,}
\def\opensquare{\ht0=3.4mm \dp0=3.4mm \wd0=6.8mm  \raise 2.7pt
\makeblankbox {.25pt} {.25pt}  }
 
 
\def\sector#1#2{\ {\scriptstyle #1}\hskip 1mm
\mathop{\opensquare}\limits_{\lower 1mm\hbox{$\scriptstyle#2$}}\hskip 1mm}
 
\def\tsector#1#2{\ {\scriptstyle #1}\hskip 1mm
\mathop{\opensquare}\limits_{\lower 1mm\hbox{$\scriptstyle#2$}}^\sim\hskip 1mm}
 
\def\p{{\bf p}}
\def\ttt{{\bf t}}
\def\uu{{\bf u}}
\def\vv{{\bf v}}
\def\ww{{\bf w}}
\def\G{{G}}
\def\S{{\bf S}}
\def\V{{V}}
\def\X{{\sst X}}
\def\Y{{\sst Y}}
\def\WS{{\sst\rm WS}}
\def\ST{{\sst\rm ST}}
\def\NS{{\sst\rm NS}}
\def\R{{\sst\rm R}}
\def\op{{\rm op}}
\def\cl{{\rm cl}}
\def\gcl{{g_\cl}}
\def\ith{i^{\rm th}}
\def\jth{j^{\rm th}}
\def\str{{\rm s}}
\def\gstr{g_{\str}}
\def\lstr{\ell_\str}
\def\eff{{\rm eff}}
\def\CFT{{\sst\rm CFT}}
%
%
\def\modulispace{1}
\def\condensation{2}
\def\Dangles{3}
\def\RGflow{4}
\def\annulus{5}
\def\cones{6}
\def\laminate{7}
\def\angledecay{8}
\def\Rcharge{9}
\def\ring{10}
\def\pseven{11}
\def\onedlattice{12}
\def\twodlattice{13}
\def\blowup{14}
\def\An{15}
\def\Anresolved{16}
\def\resolution{17}
\def\orbdecay{18}
\def\split{19}
\def\resfantwo{20}
\def\twolthree{21}

\section{Introduction}

\begin{figure}[ht]
\begin{center}
\includegraphics[scale=.6]{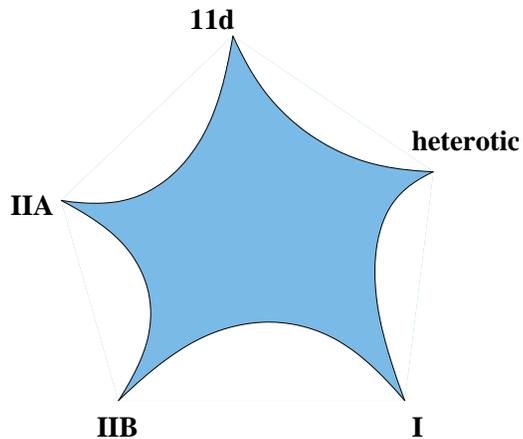}
\caption{Schematic picture of the space of string
vacua with extended supersymmetry.}
\label{modulispace}
\end{center}
\end{figure}

Vacuum selection is one of the most important issues 
in string/M theory.
The standard cartoon of string vacua (see figure \ref{modulispace})
showing all the ten dimensional perturbative string 
asymptopia together with the eleven-dimensional limit,
refers to solutions with a large number of supersymmetries,
and flat directions (moduli) in the effective action.

A much richer story arises as more supersymmetries are 
broken, and such flat directions are 
lifted or absent altogether.  
Eventually we will want to understand
vacua which respect no supersymmetries whatsoever.
To explore the configuration space, one needs to understand
off-shell physics, since the effective potential in
such situations will not be flat.

Inflation is an example of such off-shell physics
with fully broken supersymmetry;
one might consider it as a variant of `tachyon condensation'
in string theory, \ie\ the dynamical decay of an unstable
configuration.  For example, the dynamics of a D-brane
and anti D-brane, filling the noncompact dimensions of space
and separated in some compact direction, provides an
interesting model of inflation~\cite{Burgess:2001fx}.  
Here supersymmetry is fully broken, since the
half of the supersymmetries preserved on the brane
are incompatible with those preserved on the anti-brane.
Thus we are interested in understanding tachyon condensation
phenomena in string theory, for both closed and open strings.

\begin{figure}[ht]
\begin{center}
\includegraphics[scale=.5]{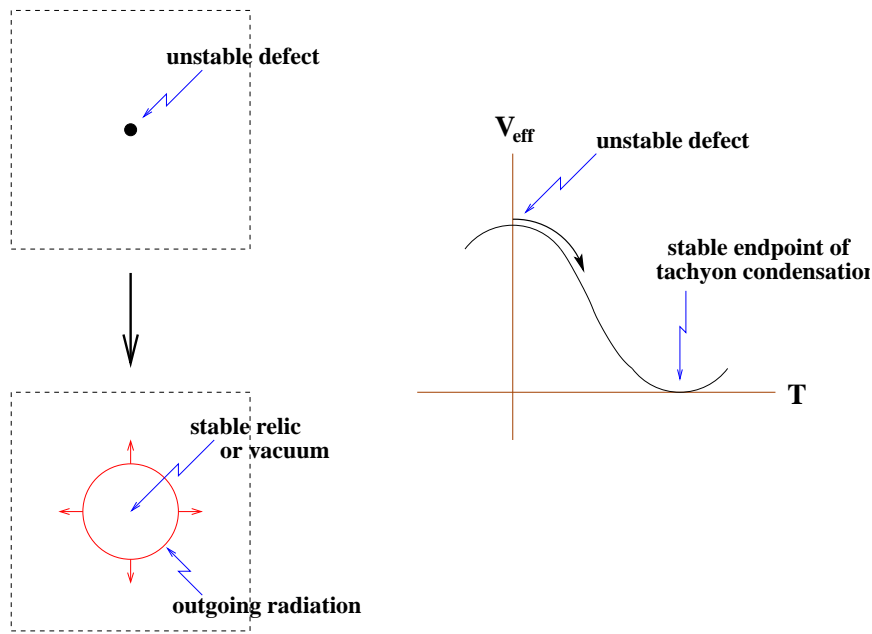}
\caption{
Decay of a localized defect leads to either
a (more) stable relic or vacuum, plus radiation.
}
\label{condensation}
\end{center}
\end{figure}

For open strings, a rather complete picture is developing,
largely due to ideas of Sen, Witten, and others%
~\cite{Sen:1999mg,Witten:1998cd}.
The presence of dynamical gravity, while certainly
of interest, is however a complication if we wish to 
isolate the condensation phenomena and their properties.%
\footnote{Consider for example the $d=26$ bosonic string.
Here condensation of the closed string tachyon might
lead to a change in the dimension of spacetime --
the only known stable vacua have $d\le 2$
(see~\cite{DiFrancesco:1995nw,Ginsparg:1993is} for reviews).}
One can eliminate this complication by studying
`impurity' or `defect' dynamics, where the tachyon
is confined to the defect, see figure \ref{condensation}.
The defect could be
\begin{itemize}
\item a D-brane system;
\item an NS5-brane system;
\item an orbifold fixed point;
\item {\etc.}
\end{itemize}
If there are sufficiently many noncompact directions
transverse to the defect, we expect to be able
to largely ignore gravitational back reaction,
or at least localize it near the defect
(and for open strings, we can tune it to zero
by taking the $\gstr\to 0$ limit).
Suppose that the bulk theory is supersymmetric, while
the defect breaks supersymmetry.  Then typically we
will have a tachyon localized at the defect,
since the ambient space far away from the defect
is locally stable.
We wish to understand the process of
condensation of this localized tachyon.

There are two common methods used to study this phenomenon.
The first is to look for time-dependent solutions of the string
equations of motion involving the growing
tachyon mode.  At asymptotically early times,
the tachyon background corresponds to a modification
of the worldsheet action describing string propagation:
\be
\SS_{\WS}\sim\SS_0+\int d^2z\, e^{EX^0}\OO_{\rm tach}
\label{timedep}
\ee
where $\OO_{\rm tach}$ is a vertex operator for the spatial 
profile of the tachyon and $X^0$ is spacetime time.
One then looks to understand the late
time behavior of the solution.  
This procedure is typically quite difficult, 
and fraught with technical difficulties associated to the fact
that little is understood about string theory in
time-dependent backgrounds -- much of the standard
treatment of string perturbation theory rests on
the connection to light cone gauge and/or analytic
continuation from Euclidean space, neither of which
are guaranteed to exist in general time dependent
backgrounds.%
\footnote{Although there is recent work
of Sen~\cite{Sen:2002nu,Sen:2002in} 
on obtaining exact solutions
for open strings via analytic continuation
from Euclidean solutions, and of 
Gutperle and Strominger~\cite{Gutperle:2002ai} on 
time-dependent solutions of the low-energy
field equations related to decaying D-branes.}

A second approach adopts
a worldsheet renormalization group (RG) viewpoint.
Conformal invariance of the worldsheet quantum field theory,
\ie\ fixed points of the 2d renormalization group,
are classical solutions of the spacetime field equations
of string theory (\cf\ \cite{Polchinski:1998rq} 
and references therein.).
Regarding the couplings $\{\lambda\}$
in the worldsheet action as
a description of the spacetime background fields,
the RG fixed point equations $\beta(\lambda)=0$
characterize background fields satisfying the classical
equations of motion, among all possible backgrounds.
For example, the usual nonlinear sigma model background
parametrized by a metric $G_{\mu\nu}$, antisymmetric tensor 
gauge field $B_{\mu\nu}$, and dilaton $\Phi$,
yields the usual low-energy field equations
\bbb
\beta_{\mu\nu}^{\sst (G)} &=&
	\alpha'\RR_{\mu\nu}+2\alpha'\nabla_\mu\nabla_\nu\Phi
	-\coeff14{\alpha'} H_{\mu\lambda\kappa}H_\nu^{~\lambda\kappa}
	+O({\alpha'}^2) 
\nonumber\\
\beta_{\mu\nu}^{\sst(B)} &=&
	-\coeff12 \alpha'\nabla^\lambda H_{\lambda\mu\nu}
	+\alpha' \nabla^\lambda\Phi H_{\lambda\mu\nu}
	+O({\alpha'}^2)
\label{betafneqs}\\
\beta^{\sst(\Phi)} &=&
	\coeff{D-10}4 - \coeff12\alpha'\nabla^2\Phi
	+\alpha'\nabla_\lambda\Phi\nabla^\lambda\Phi
	-\coeff1{24}\alpha'H_{\lambda\mu\nu}H^{\lambda\mu\nu}
	+O({\alpha'}^2)
\nonumber
\eee
as the condition of conformal invariance; similarly,
scale invariance under boundary perturbations yields
the open string equations of motion%
~\cite{Fradkin:1985qd,Tseytlin:1986ti,Metsaev:1987qp,Abouelsaood:1987gd}.

Nonconformal backgrounds in the worldsheet
field theory provide a way of continuing
off-shell; RG flows interpolate between classical
solutions, and thus provide information about the topology
of the effective action and the configuration space.

The mass shell condition
$L_0-1/2 = \alpha'(-E^2+\p^2)/2+h_{\rm int}=0$,
where $h_{\rm int}$ is the contribution to the conformal dimension
of a vertex operator coming from internal structure
(\ie\ the conformal field theory (CFT) describing the defect), gives
the correspondence
%
\be
\alpha'm^2 
= (h_{\rm int}-\hf)\ 
\cases{
	{>0}	&	${{\rm massive~in~spacetime\hfill}
			\atop{\rm irrelevant~on~worldsheet}\hfill}$\cr
	& \cr
	{=0}	&	${{\rm massless~in~spacetime\hfill}
			\atop{\rm (nearly)~marginal~on~worldsheet\hfill}}$\cr
	& \cr
	{<0}	&	${{\rm tachyonic~in~spacetime\hfill}
			\atop{\rm relevant~on~worldsheet\hfill}}$
}
\ee
between spacetime properties and the effect under the
renormalization group of perturbing the worldsheet action 
by the vertex operator.

Tachyon condensation thus corresponds to adding a relevant
operator to the worldsheet Lagrangian describing the
background in which perturbative strings propagate
\be
\SS_{\WS}=\SS_0+\lambda(t)\int\! d^2z\,\OO_{\rm tach}\ .
\ee
This breaks conformal symmetry; the perturbation grows toward
the IR on the worldsheet, and the endpoint of tachyon
condensation in this context is the IR fixed point of the 
worldsheet renormalization group flow.
Near the UV fixed point (where the tachyon perturbation
vanishes), the coupling scales as
\be
\lambda(t)\sim e^{E^2 t}
\label{scaledep}
\ee
where $t$ is the (logarithmic) worldsheet scale.
Note that there is not a direct correspondence between
spacetime time dependence \pref{timedep} 
and worldsheet scale dependence \pref{scaledep};
in particular the effective spacetime equations of motion are
second order differential equations, while the RG equation
is first order.  Nevertheless, RG flows define interesting
paths in the configuration space leading away
from unstable extrema toward more stable ones; 
and there is no known dynamical decay process for
which there is no corresponding RG flow.

One can also relate the renormalization group 
more closely to spacetime dynamics via light front evolution%
~\cite{Tseytlin:1993pq}.
Given an RG flow in the nonlinear sigma model
with $d$ dimensional target space, one can make a
conformally invariant theory in $D=d+2$ dimensional spacetime via
\bbb
ds^2	&=&	-2dudv+g_{ij}(u,x)dx^idx^j
	\nonumber\\
\Phi	&=&	pv+\phi(u,x)
\eee
with $p={\it const.}$;
for the full set of conditions on the fields,
the reader is referred to%
~\cite{Tseytlin:1993pq}.
The salient feature is that the equation of motion 
\pref{betafneqs} for the transverse metric reduces to
\be
p\frac{\partial g_{ij}}{\partial u} = \beta^{(g)}_{ij}
	+D_{(i}W_{j)}+2D_iD_j\phi\ ,
\ee
so that one can interpret the null coordinate $u$
as `RG time'.  The RG flow becomes the profile of
a gravitational wave in spacetime, and the geometry
interpolates between fixed points of the RG flow
of the $d$-dimensional transverse geometry
as one traverses the wave.

In these lectures, we will explore the renormalization
group approach to the condensation of tachyons 
on localized defects in string theory,
following~\cite{Harvey:2000na,Harvey:2001wm}.
In section 2 we review a variety of results from
the study of open string tachyon condensation on
collections of D-branes, which may be simply
obtained using the RG approach.  The RG trajectories are specified
by the gradient of the open string effective action,
which therefore monotonically decreases along the flows.
We explain the connection of this effective action
to the density of states and the intuition that the
renormalization group is a process of thinning of degrees 
of freedom (the `dissipated states' of the title).  
We then explore in section 3
the extent to which these ideas can be carried over
to closed string tachyon condensation, using the
example of nonsupersymmetric orbifolds as the prototype
of an unstable closed string defect.  The examples studied
exhibit rich connections to algebraic geometry,
which we exploit to map out the set of RG flows.
An attempt is made to characterize the flows
by a decrease of the density of states associated
to the defect, analogous to the open string case,
with inconclusive results.


\section{Open strings}

For open strings, one has {\it boundary RG flow}
\be
\SS_{\WS}=\SS_0+\lambda\int_\partial ds\,\OO_{\rm tach}\ .
\ee
The worldsheet bulk theory remains conformal;
in this case the RG flow interpolates between two different
boundary states of the same target space --
in other words, the defect decays.

In this section we survey open string RG flows,
restricting our attention to the superstring 
unless otherwise indicated.
The examples we will consider are:
\begin{itemize}
\item
$Dp$-$\overline{Dp}\longrightarrow{\it vacuum}$
\item
$Dp$-$\overline{Dp}\longrightarrow D(p-2)$
\item
$Dp$-$\overline{D(p-2)}\longrightarrow Dp~{\it with~magnetic~flux}$
\item
$N$ $D0$ branes on $SU(2)$ $\longrightarrow$ $D2$ 
\end{itemize}
The first of these examples is the complete decay of an unstable
brane system.  In the remaining examples, conserved
topological charges prevent the complete decay of the 
initial brane configuration.  These topological charges
are the subject of K-theory, which is roughly the classification
of vector bundles up to isomorphism -- appropriate
since open string dynamics is that of (generalized) gauge fields.
The last decay ($D0\to D2$) is an example of the Myers effect%
~\cite{Myers:1999ps};
in the context of the renormalization group, one finds
an amusing application of one of its earliest successes,
namely the Kondo model%
~\cite{Wilson:1975mb} (for a review, see~\cite{Affleck:1995ge}).

\begin{figure}[ht]
\begin{center}
\includegraphics[scale=.6]{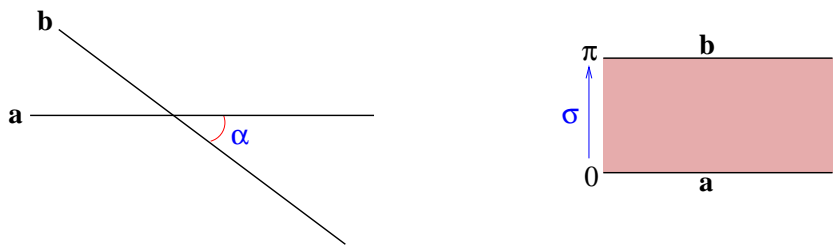}
\caption{
A pair of D-branes at relative angles has an
open string tachyon localized along their intersection.
The tachyon lives in the sector of open strings
with mixed $a$--$b$ boundary conditions depicted
on the right.
}
\label{Dangles}
\end{center}
\end{figure}

A basic property that we will be using is the spectrum
of open strings connecting two D-branes (label them
$a$ and $b$) oriented at an angle $\alpha$ 
relative to one another (see figure \ref{Dangles}).  
The boundary conditions on such strings are
\bbb
{\it Re}\,\partial_\sigma X|_{\sigma=0}=0\quad&,&\quad
{\it Re}\,e^{i\alpha}\partial_\sigma X|_{\sigma=\pi}=0
\nonumber\\
{\it Im}\,X|_{\sigma=0}=0\quad&,&\quad
{\it Im}\,e^{i\alpha}X|_{\sigma=\pi}=0
\eee
and similarly for the worldsheet superpartners $\psi$ of $X$.

Solving for the string oscillator spectrum, one finds the moding
shifted by $\alpha/\pi$.  This is just like the twisted sectors
of $\IZ_N$ orbifolds; thus one deduces that there 
exists an operator
\be
\Sigma_\alpha=\sigma_\alpha \,\exp[i(\alpha/\pi)H]
\ee
(where $e^{iH}=\psi_1+i\psi_2$ bosonizes the worldsheet fermions,
and $\sigma_\alpha$ implements the twisting of $X$),
which creates the lowest mass $a$--$b$ open string when
acting on (say) the $a$--$a$ ground state.%
\footnote{The conjugate operator $\Sigma^*$ creates the
conjugate string when acting on the $b$--$b$ open string.}
The calculation of the conformal dimension of this
open string twist field will be exactly as for
the corresponding closed string twist operator%
~\cite{Dixon:1987qv,Hamidi:1987vh,Bershadsky:1987fv};
one finds
\be
h_\Sigma = \half\frac{\alpha}{\pi}\Bigl(1-\frac{\alpha}{\pi}\Bigr)
		+\half\Bigl(\frac{\alpha}{\pi}\Bigr)^2
         = \half\frac\alpha\pi
\ee
($0<\alpha<\pi$).  This operator is GSO odd; one gets a GSO even
operator by applying $\psi^*=e^{-iH}$ to get
\be
\tilde\Sigma_\alpha = \sigma_\alpha\, \exp[-i(1-\alpha/\pi)H]
\ee
whose conformal dimension is $h_{\tilde\Sigma}=\hf(1-\frac{\alpha}{\pi})$.
Thus there is a tachyon whenever the angle between
the branes is nonzero.

Among the interesting values of $\alpha$ are
\begin{itemize}
\item
$\alpha=0$ describes two BPS $Dp$ branes;
\item
$\alpha=\pi$ describes $Dp$-$\overline{Dp}$;
\item
$\alpha=\pi/2$ is related via T-duality to $Dp$-$D(p-2)$.
\end{itemize}
In the last two cases there is a tachyon in the spectrum
related to the decays listed above.
Let us discuss their condensation.


\subsection{$Dp$-$\overline{Dp}$ $\longrightarrow$ {vacuum}}

Consider first $\alpha=\pi$.  One expects that the branes
can and will decay to vacuum plus radiation, for instance
the $D0$-$\overline{D0}$ system is a pair of 11d gravitons
with opposite momentum on a circle in M-theory, 
which can scatter into final states with no momentum
on that circle.  This expectation of decay has been verified
by a variety of complementary approaches:
\begin{itemize}
\item
String field theory~\cite{Sen:1999nx};
\item
Noncommutative effective field theory~\cite{Harvey:2000jt};
\item
Boundary effective field theory (BEFT)/RG flow%
~\cite{Gerasimov:2000zp,Kutasov:2000qp,Kutasov:2000aq}.
\end{itemize}

The BEFT/RG approach%
\footnote{Sometimes called boundary string field theory (BSFT).
This is somewhat of a misnomer; since the approach uses
continuum Euclidean worldsheet field theory, it is restricted to 
massless and tachyonic perturbations%
~\cite{Banks:1987qs},
and so is not a field theory of the entire string spectrum.}
relies on the following property of the disk worldsheet path integral
~\cite{Tseytlin:1988ww,Andreev:1988cb,%
Gerasimov:2000zp,Kutasov:2000qp,Kutasov:2000aq,Niarchos:2001si}
\be
\Gamma_{\ST}(\lambda_i)
	=\ZZ_{\WS}(\lambda_i)
	=\int\DD X\DD\psi\,e^{\SS_0}\;\Tr\,
		\exp\Bigl[-\int_{\partial}dsd\theta\,\lambda_i\OO^i\Bigr]\ .
\ee
In other words, the conditions of vanishing
of the beta functions which are the spacetime 
equations of motion, can be expressed as
\be
\frac{\partial\Gamma_{\ST}}{\partial \lambda_i}
	= \frac{\partial\ZZ{\WS}}{\partial \lambda_i}
	= \beta^j(\lambda)G_{ij}
\ee
where $G_{ij}$ is the (Zamolodchikov) metric on the coupling space.
A further important property is that the variation 
of the disk partition function with respect to worldsheet scale is
\be
\frac{\partial\Gamma_{\ST}}{\partial t}
	= -\beta^i\frac\partial{\partial \lambda_i}\Gamma_{\ST}
	= -\beta^i\beta^jG_{ij}\ ;
\ee
for unitary worldsheet quantum field theories
the RG flow is monotonically in the direction of decreasing
spacetime action, since the metric $G_{ij}$ is positive definite.
Thus flows by relevant operators agree with the idea
that tachyon condensation is a process of rolling down
the effective potential to a more stable extremum%
~\cite{Gava:1997jt,Elitzur:1998va,Harvey:2000na}.

The relation $\Gamma_{\ST}=\ZZ_{\WS}$
is illustrated by the example of open string
gauge fields.  One finds the Dirac-Born-Infeld action%
~\cite{Fradkin:1985qd,Abouelsaood:1987gd}
as the spacetime effective action for smoothly varying
gauge fields
\be
\Gamma_{\ST}=\mu_p\int\! d^{p+1}x\, e^{-\Phi}
	\sqrt{\det[G+B+2\pi\alpha' F]}
\ee
from a computation of the disk partition function
(here $\mu_p$ is the brane tension).
More generally, one has a perturbation
\be
\lambda_i\OO^i = \bar\Gamma D\Gamma +\Gamma\cdot T(X)
	+\bar\Gamma\cdot\bar T(X) + A_\mu DX^\mu +\ldots
\ee
with $D=\partial_\theta+\theta \partial_s$ is the 
derivative on boundary superspace, and $\Gamma=\eta+\theta f$
is an auxiliary fermionic superfield living only on the boundary;
the latter is needed by the Grassmann parity of the boundary integration
measure, so that the tachyon action is Grassmann even.
If there are $k$ such superfields, their 
$2^k$ dimensional Hilbert space is a convenient way to
realize a Chan-Paton vector space.  A calculation of
the effective action yields%
~\cite{Kutasov:2000aq,Kraus:2000nj,Takayanagi:2000rz}
\be
\Gamma_{\sst Dp-\overline{Dp}} =
	2\mu_p \int\! d^{p+1}\!x\, e^{-\Phi}e^{-2\pi\alpha'|T|^2}
	\Bigl[1+8\pi\alpha'\log2|D_\mu T|^2+\coeff{2\pi\alpha'}{8}
		(F_{\sst Dp}^2+F_{\sst\overline Dp}^2)+\ldots\Bigr]\ .
\ee
The overall factor of $\exp[-2\pi\alpha'|T|^2]$ is in hindsight
very easy to understand; if one perturbs by constant $T$, this 
has no effect on the worldsheet dynamics, but the partition
function rescales as (upon integrating the boundary action
over the boundary supercoordinate, 
and suppressing the $\eta$ integral which
merely constructs the Chan-Paton bundle)
\bbb
\ZZ_{\rm disk} &=& \int \DD X\DD\psi\DD\! f \;
	e^{-\SS_0}\, \exp\Bigl[\coeff{1}{2\pi\alpha'}
		\int_{\partial} ds\,[fT+\bar f\bar T+|f|^2]\Bigr]
\nonumber\\
	&=& \Bigl(\exp[-2\pi\alpha'|T|^2]\Bigr)\;
		\int \DD X\DD\psi\, e^{-S_0}\ .
\eee
Condensation of $T$ leads to suppression of all disk amplitudes,
\ie\ open strings disappear from the dynamics as the $D$-$\bar D$
system decays away.  The corresponding IR fixed point of the
renormalization group is trivial.

The relation $\Gamma_{\ST}=\ZZ_{\WS}$
is a universal property of open string theory on which 
the BEFT approach relies; the proof of%
~\cite{Kutasov:2000qp,Marino:2001qc,Niarchos:2001si}
relies only on general properties of the worldsheet RG
such as the Callan-Symanzik equation.
This relation is possible because of the nature of the M\"obius
volume of the disk%
~\cite{Tseytlin:1988ww,Andreev:1988cb,Liu:1988nz}
of the form ${\it Vol}(\mbox{\sl PSL}(2,\IR)=\coeff{c_1}\epsilon+c_2$
with $c_{1,2}$ finite constants, 
leaving a finite remainder after subtraction;
even better, $c_1=0$ for the superstring due to Bose-Fermi cancellations.
It is believed that a similar approch should work for
the closed string, however there are some unresolved puzzles.
The corresponding M\"obius volume 
${\it Vol}(\mbox{\sl PSL}(2,\IC))=\coeff{c_3}{\epsilon^2}+c_2\log\epsilon$;
clearly the finite term is not universal.
There is a suggestion~\cite{Tseytlin:1988tv} 
that the spacetime action is 
$\partial\ZZ_{\sst \rm sphere}/\partial\log\epsilon$,
which gives the right answer to one-loop order
in the nonlinear sigma model on compact targets.
It would be interesting to know if this idea or
something similar could define the effective action
$\Gamma_{\sst \rm ST}^{\sst\rm closed}$ in terms of quantities
intrinsic to the worldsheet field theory.


\subsection{{$Dp$-$\overline{Dp} \longrightarrow D(p-2)$}}

\begin{figure}[ht]
\begin{center}
\includegraphics[scale=.6]{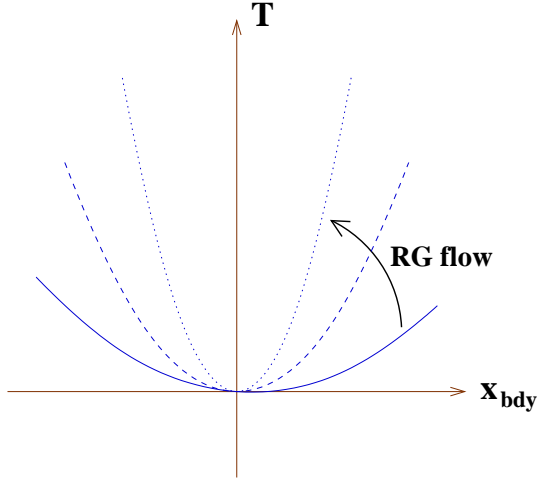}
\caption{
A boundary potential deepens under RG flow, confining
boundary coordinate(s) to a lower dimensional subspace.
}
\label{RGflow}
\end{center}
\end{figure}

Consider the boundary perturbation
\be
\delta\SS=\lambda\int \!dsd\theta\,(\Gamma X+\bar\Gamma \bar X)\ ,
\ee
where $X=X^1+iX^2$ are coordinates along the brane;
after eliminating the auxiliary field $f$, this is a 
generalized mass term.  
In the Hilbert space of the boundary fermion $\eta$,
the state with fermion number zero (one) is
the (anti) brane.
The worldsheet bulk remains conformal under such a perturbation,
but the string coordinates have a mass on the boundary that
breaks worldsheet conformal invariance and also
target space translation symmetry.  
The formerly free boundary value of $X$ 
becomes confined to $X=0$ in the worldsheet IR,
where $\lambda(t)\to\infty$ (see figure \ref{RGflow}).  
The effective boundary condition
passes from Neumann to Dirichlet along two directions 
parallel to the brane, and $Dp$-$\overline{Dp}\longrightarrow D(p-2)$.
The winding of the phase of the tachyon at spatial infinity
is the $D(p-2)$ charge (so that $T\sim x^n$
makes $n$ lower branes).

One can generalize this construction%
~\cite{Witten:1998cd,Horava:1998jy}
to other mass terms by adding indices
\be
\delta\SS=\lambda\int\! dsd\theta\,(\Gamma_i X^i+\bar\Gamma_i \bar X^i)
\ee
to get higher codimension branes at the infrared fixed point
(\cf~\cite{Kutasov:2000aq});
$\Gamma\cdot X$ realizes a coupling of orientation in spacetime
and Chan-Paton vector bundle (realized by the Hilbert space
of the boundary fermions $\eta$).  A more elaborate example,
that of $D0$ branes on ${\S}^3$, will be discussed below.

Often, the endpoint of tachyon condensation is the lightest object
carrying given topological (RR) charges, indeed there is a close
connection between RR charge and K-theory~\cite{Minasian:1997mm}.
The coupling of D-branes to RR charge can be seen in the boundary
effective field theory formalism by computing $\ZZ_{\rm disk}$
with the insertion of a RR vertex operator in the center
of the disk (see for instance~\cite{Kraus:2000nj}).
One finds
\be
\ZZ_{\rm RR} = \mu_p \int C\wedge {\rm STr}\,\exp[2\pi i\alpha'\FF]
	\sqrt{\hat\AA(R)}\ .
\label{ZRR}
\ee
The trace {\it STr} is graded so that branes count with a plus
sign and antibranes with a minus sign (\ie\ $\Tr[(-1)^F\cdots]$
in the Hilbert space of the boundary fermions).
Here $\FF$ is the curvature
\be
i\FF = \pmatrix{iF_{Dp}-T\bar T & \DD\bar T\cr
		 \DD T & iF_{\overline{Dp}}-\bar T T}
\ee
of a generalized connection
\be
\AA = \pmatrix{iA_{Dp} & \bar T \cr
		T & iA_{\overline{Dp}}}
\ee
known as the {\it Quillen superconnection}.
The idea is that the boundary fermion $\Gamma=\eta+\ldots$
allows us to think of $T\cdot\Gamma$ as a one form on
the Chan-Paton vector space, just as $A_\mu DX^\mu$
codes one forms in target space via the boundary value of
the fermion $DX=\psi+\ldots$.  
Expanding the generalized Chern character
\pref{ZRR}, there is a term
\be
\mu_p\int C_{p-2}\wedge e^{-2\pi\alpha' |T|^2}
	(2\pi\alpha')^2 dT\wedge d\bar T
\ee
that gives the coupling to the RR ($p-2$)-form, 
indicating that the field configuration on the branes
indeed properly carries $D(p-2)$ brane charge.


\subsection{$Dp$-$D$($p$-2)}

The third  example is the tachyon appearing
between $Dp$ and $D(p-2)$ branes.
Because of the coupling $\int C_{p-2}\wedge F_{\sst Dp}$ 
in the expansion of \pref{ZRR},
a $Dp$ brane can itself carry $D(p-2)$ brane charge.
Let the directions parallel to the $Dp$ but orthogonal
to the $D(p-2)$ be compactified on a volume $V_2$.
The action of the $Dp$ brane is
\be
\Gamma_{\sst\rm DBI} =
	\mu_p\int d^p\!x\,e^{-\Phi}\sqrt{\det[G+B+2\pi\alpha'F]}\ ;
\ee
with $n_{p-2}$ units of $F$ flux on the compact volume $V_2$,
the action of the minimal energy configuration reduces to
\be
\mu_p\sqrt{n_p^2V_2^2+n_{p-2}^2(2\pi\alpha')^2}
	\int d^{\sst p-2}\!x\,(\cdots)
\label{branebound}
\ee
which is smaller than the action per unit $(p-2)$ volume,
$(n_pV_2+n_{p-2}2\pi\alpha')\mu_p$,
of the UV fixed point configuration
of $n_p$ $Dp$-branes and $n_{p-2}$ $D(p-2)$-branes separately.
This is consistent with the idea that the configuration
of branes with flux is the endpoint of tachyon condensation;
a calculation demonstrating that this is indeed the endpoint
was carried out by Gava, Narain and Sarmadi%
~\cite{Gava:1997jt}.

The fact that a $Dp$ brane with $F$ flux is a $Dp$-$D(p-2)$
bound state means that another way to make the $D(p-2)$ brane
is to condense the tachyon between this bound state and a
$\overline{Dp}$; the higher branes will disappear, leaving
the lower brane behind.


\subsection{Systems with less supersymmetry}

So far, we have considered just the simplest 
properties of branes in situations with a large amount
of supersymmetry.  One would like to know how much of
the above story generalizes to situations with less 
supersymmetry, say $\NN=1$ in four dimensions.
Douglas and collaborators have made much progress
in this direction (\cf~\cite{Douglas:2001hw} and 
references therein)
in examples with $\NN=2$ worldsheet supersymmetry.
This amount of worldsheet supersymmetry 
in the absence of D-branes implies
$\NN=2$ spacetime supersymmetry in type II string
theory without branes~\cite{Dixon:1987bg,Banks:1988yz};
half-supersymmetric brane configurations then break
the symmetry further to $\NN=1$ in spacetime.

To begin, let us review the connection
between worldsheet and target supersymmetry.
Consider a worldsheet CFT with $\NN=(2,2)$
supersymmetry; one has the supermultiplet
of currents containing the stress tensor
$T$; two supercurrents $G$, $G^*$; and the
$U(1)$ R-current $J$ that rotates the supersymmetries
by opposite phases (and similarly for the right-movers,
hence the doubled notation $\NN=(2,2)$).
A typical example is type II on a Calabi-Yau $n$-fold;
there $J$ is related to the $U(1)$ holonomy
preserved by the target in the geometrical
(large volume) limit.
The spacetime $\NN=2$ supercharges are related to
the integrals of worldsheet supercurrents, whose only
dependence on the $\NN=(2,2)$ worldsheet theory is
through the operator that implements {\it spectral flow}
between the NS and R sectors.  In any worldsheet theory
with a $U(1)$ current algebra symmetry, one can
decompose all the vertex operators $\OO$ 
that are highest weight under the $J$ current algebra
into the form
\be
\OO=\Psi\, e^{iqH}\ ,
\label{pfrep}
\ee
where $H$ bosonizes the current via $J=iQ_0\partial H$,
and $\Psi$ is an operator that commutes with $J$.  
In other words, the exponential carries the $U(1)$ charge dependence.
Using conventional normalization for the scalar field $H$,
$\vev{H(z)H(0)}=-\log z$, the two-point function 
of the current 
\be
\vev{J(z)J(0)} = \frac{[\half(10-d)]^{1/2}}{z^{2}}
\ee
determines $Q_0=[{(10-d)/2}]^{1/2}$
for the $\NN=2$ CFT describing compactification down
to $d$ dimensions.
One can then represent the spacetime supersymmetry
charges as
\bbb
\QQ &=& \oint dz \,({\bf q}_{\IR^{d-1,1}})\;\exp[i(Q_0/2)H]
\nonumber\\
\tilde\QQ &=& \oint\bar dz\,(\bar {\bf q}_{\IR^{d-1,1}})\;
	\exp[i(Q_0/2)\bar H] 
\eee
where ${\bf q}_{\IR^{d-1,1}}$ is a holomorphic operator
of worldsheet dimension $h=\coeff1{16}(6+d)$ giving the
contribution to the spacetime
supersymmetry operator coming from the noncompact directions
and the worldsheet BRST ghosts
(\cf~\cite{Polchinski:1998rr}).

BPS D-branes will break the spacetime supersymmetry to $\NN=1$
via the worldsheet boundary conditions%
~\cite{Ooguri:1996ck}
\be
\exp[{\coeff i2 Q_0H}]=\exp[{i\alpha}]\,\exp[\coeff i2 Q_0\bar H]\ .
\ee
The phase $\alpha$ determines which of the $\S^1$ family of
$\NN=1$ subalgebras of the spacetime $\NN=2$ supersymmetry
is preserved by the brane being described by the given boundary
condition; the key point is that not all branes have
the same $\alpha$ -- supersymmetry is broken for a pair
of branes $a$, $b$ with $\alpha_a\ne\alpha_b$.

The Ramond sector ground states in the $a$--$b$ sector of
such a pair will still be chiral fermions in spacetime
and therefore massless (the GSO projection acts as a chirality
projection on Ramond ground states).  
Furthermore, spectral flow still relates NS and R boundary conditions,
since we still have $\NN=2$ worldsheet supersymmetry.  Thus
\bbb
\OO_{ab}^{\NS} &=& \Psi\,\exp\Bigl[i\frac{q}{Q_0}H\Bigr] 
\nonumber\\
\OO_{ab}^{\R} &=& 
	\Psi\, \exp\Bigl[i\Bigl(\frac{q}{Q_0}-\frac{Q_0}2\Bigr)H\Bigr]
	\;({\bf q\bar q})_{\IR^{d-1,1}} \ .
\eee
The dimension $h_{ab}^{\R}=1$ of $\OO^\R_{ab}$ is determined
by the masslessness of the Ramond ground state;
as above, this gets a contribution $\coeff1{16}(6+d)$ from
the operator $({\bf q\bar q})_{\IR^{d-1,1}}$ involving the noncompact
directions and worldsheet BRS ghosts, leaving a
contribution $Q_0^2/8$ from the compactification CFT.
Therefore we can deduce the dimension of $\Psi$ to be
\be
h_\Psi=\frac{Q_0^2}8-\half\Bigl(\frac q{Q_0}-\frac{Q_0}2\Bigr)^2\ ,
\ee
and thus the mass of the NS ground state operator
$\OO_{ab}^{\NS}$ is
\bbb
m^2=h_{ab}^{\NS} -\hf
	&=& \frac{Q_0^2}8-\half\Bigl(\frac q{Q_0}-\frac{Q_0}2\Bigr)^2
		+\half\Bigl(\frac{q}{Q_0}\Bigr)^2
\nonumber\\
	&=&\hf(q-1)\ .
\eee
Here $q=(\alpha_a-\alpha_b)/\pi$.  

In fact, we saw a simple
example of this sort of calculation before, in the
context of two D-branes at relative angles; the mismatch
of preserved supersymmetries there gives a direct interpretation
to $q$ as the relative angle $\alpha$ between the branes.
One way of generating the angle $\alpha$ is to wrap two
branes on the $a$ and $b$ cycles of a two-torus $T^2$;
then the angle between the branes depends on the modulus $\tau$
of the torus, $\alpha=\alpha(\tau)$.  This T-dualizes
to $\alpha=\alpha(B+iV_2)$ and branes wrapping even-dimensional cycles,
\cf\ equation \pref{branebound}.
The generalization to higher dimensions and less supersymmetry
is a dependence on the complexified K\"ahler moduli of the spectrum on
such branes 
$\alpha_{ab}=\alpha_{ab}(\MM_{{\sst\rm K\ddot ahler}})$.
A property of branes on the torus is $\alpha_{ab}<0$; this need not
be true in the general case (\ie\ there need not always be
a tachyonic $a$--$b$ string whose condensation generates
a bound state of branes $a$ and $b$).

The Quillen superconnection mentioned above in a sense
generalizes to a differential on the whole complex of 
potential brane bound states carrying BPS charge vectors.
This complex is called the {\it derived category}.
One has again an equivalence relation between objects 
whose physical origin is the possibility to add brane-antibrane
pairs and condense the resulting tachyons
(recall that the tachyon is the `off-diagonal'
part of the superconnection connecting different objects);
objects modulo equivalence then classifies
D-brane charges in a way that generalizes K-theory.
For instance, mirror symmetry is automatically incorporated
since this operation is simply T-duality 
($\bar J\to -\bar J$, $\bar H\to -\bar H$)
on the $U(1)$ R-current of the worldsheet $\NN=2$.
Also, the limit of stringy target spaces,
whose geometrical interpretation is more obscure,
is incorporated as well.


\subsection{D0's on SU(2): the Kondo model}

The final example of open string tachyon condensation
that we will consider is that of $D0$ branes in the
$SU(2)$ WZW model, \ie\ the three-sphere $\S^3$ threaded
by $k$ units of NS three-form flux $H$%
~\cite{Bachas:2000ik,Alekseev:2000fd}.
Suppose there are $N$ coincident $D0$ branes on the $\S^3$.
Then there is a marginally relevant (\ie\ logarithmically
growing toward the IR) boundary perturbation
\be
\delta\SS = \int_\partial ds\, j^a(s)S_a\ ,
\ee
where the $S_a$ are the representation matrices of $SU(2)$
in the $N$-dimensional representation, acting on the
Chan-Paton Hilbert space of the branes.
The worldsheet theory is none other than the {\it Kondo model} --
currents interacting with a localized impurity in 2d,
taking us back to the very origins of the renormalization group%
~\cite{Wilson:1975mb}.

The total spin $j^a+S^a$ is conserved by the interaction,
so (at least for $N\ll k$) one expects the infrared fixed point
to be {\it some} boundary state in $SU(2)$ CFT which respects
the global $SU(2)$ symmetry.  Boundary states of this sort,
having the boundary condition $j_a=\bar j_a$ on the currents,
are in one-to-one correspondence 
with integral conjugacy classes of the group
(\cf~\cite{Elitzur:2000pq} for a pedagogical discussion), 
so there is one for each $SU(2)$ spin $j=(N-1)/2=0,\hf,1,\dots$.
These representations can be seen 
to describe spherical $D2$ branes of size
\be
R_{S^2} = \sqrt k\,\sin\Bigl[\frac{\pi N}{k}\Bigr]\ .
\ee
Here tachyon condensation results in the `dielectric effect'%
~\cite{Myers:1999ps},
where $N$ $D0$ branes in the presence of an $H$ field puff up
into a $D2$ brane.


\subsection{The $g$-theorem}

As mentioned above, the relation $\Gamma_{\ST}=\ZZ_{\WS}$
gives a purely worldsheet criterion for allowed decays --
namely, the disk partition function must decrease
along the corresponding RG flow
\be
\Gamma_{\sst\rm IR}<\Gamma_{\sst\rm UV}\ .
\ee
The existence of a function on open string 
coupling space that decreases along RG flows 
is known as the {\it $g$-theorem};
it was conjectured by Affleck and Ludwig%
~\cite{Affleck:1991tk}
(who called $\Gamma_{\ST}\equiv g$)
and proven by Kutasov, Marino and Moore%
~\cite{Kutasov:2000qp}.
As a corollary, we deduce that $\Gamma=g$ is constant
along marginal lines of boundary perturbations.%
\footnote{Note that $\Gamma$ need {\it not} be constant
under deformations by closed string moduli; for instance,
the spacetime energy of a D-brane wrapped on a
torus depends on the volume of the torus.}

\begin{figure}[ht]
\begin{center}
\includegraphics[scale=.5]{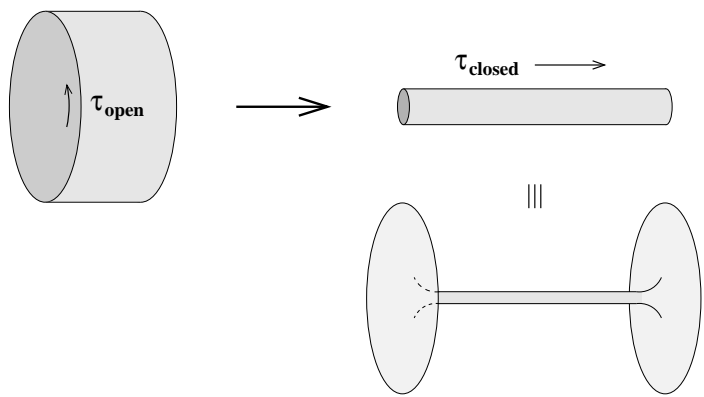}
\caption{
The annulus partition function factorizes 
as $\tau_{\sst\rm open}\to 0$ onto the square
of the disk partition function.
}
\label{annulus}
\end{center}
\end{figure}

By the magic of worldsheet channel duality, there is
another interpretation of $\Gamma=g$.  Consider the
annulus partition function in the limit of vanishing
open string proper time of propagation $\tau_{\sst\rm open}\to 0$
(see figure \ref{annulus})
\be
\ZZ_{\sst\rm annulus} = \Tr_{\HH_{aa}}\,
	e^{2\pi i\tau_{\sst\rm open}(L_0-c/24)}
\ee
By modular invariance, the amplitude factorizes in
the closed string channel on the lightest closed string
state, namely the identity operator, leaving two copies
of the disk partition function.

\noindent
Thus
\be 
\lim_{\tau_{\sst\rm open}\to 0} \,\Tr_{\HH_{aa}}\,
	e^{2\pi i\tau_{\sst\rm open}(L_0-c/24)}
	=e^{2\pi i\tau_{\rm closed}(-c/24)}
		\Bigl(\Gamma_\ST^{\sst(a)}\Bigr)^2\ ,
\ee
with $\tau_{\rm closed}=-1/\tau_{\rm open}$.
On the other hand, the LHS is dominated by the undamped exponential
degeneracy of the high energy spectrum of states in the open
string Hilbert space; converting the sum to an integral
over the density of states, $\Tr_{\HH_{aa}}\to\int dE\,\rho_{\rm open}(E)$,
where $E=L_0-c/24$, performing an inverse Laplace transform
in $\tau_{\rm open}$ and making a saddle point approximation to
the integral, one finds
\be
\rho_{\rm open}(E)=
	\half \Gamma_{\sst(a)}^2\Bigl(\frac{c}{6 E^3}\Bigr)^{1/4}
		\exp\left[2\pi\sqrt{cE/6}\right]\ .
\label{rhoopen}
\ee
Now $c$ is a property of the bulk of the worldsheet CFT, and cannot change
under boundary RG flow (for instance, $c$ can be measured as the
leading coefficient in the stress tensor OPE at points well away
from the boundary).  Thus another way to state the $g$-theorem
is that the aymptotic density of open string states
monotonically decreases under boundary RG flow,
which is to say open string tachyon condensation.
This is an embodiment of the usual intuition that the
renormalization group implements a `thinning'
of field theoretic degrees of freedom.


\section{Closed strings}

We are now interested in asking the analogous questions for
closed strings.  Adopting the worldsheet renormalization group
approach to tachyon condensation, we would like to know
\begin{itemize}
\item
{\bf Q:}
What is the structure of RG flows for localized nonsupersymmetric defects?\\
{\bf A:} We will find that, as in the open string case,
$\NN=(2,2)$ worldsheet supersymmetry is a powerful tool,
enabling us to understand the flows in a variety of examples.
\item
{\bf Q:} Is there an analogue of the $g$-theorem, in the sense of 
(a) a decrease of the closed string effective action along flows; or 
(b) a decrease in the asymptotic density of states?\\
{\bf A:} There is no obvious analogue of (a)
for the closed string case; we will construct the obvious candidate
for (b), and show that it does decrease in a wide variety of
examples.
\item
{\bf Q:} Is there an analogue of K-theory or the derived category?\\
{\bf A:} Who knows?
\end{itemize}

A concrete class of examples of localized closed string tachyons
arises in noncompact orbifolds of the type
$\IR^{d-1,1}\times(\IR^{10-d}/\G)$, where the orbifold
group $G\subset SO(10-d)$ but $G\not\subset SU(5-d/2)$.
Then supersymmetry is broken by the boundary conditions in
twisted sectors of the orbifold, but not in the untwisted
sector if we perform the usual GSO projection that gives type IIA/B
string theory.  The twisted sectors indeed contain closed string
tachyons; where do their RG flows take the theory?

The first question to answer is why the central charge does
not decrease, leading us to noncritical string theory;
in other words, why doesn't $c$ play the role of $g_{\sst\rm closed}$,
and why doesn't space just disappear?  After all, Zamolodchikov%
~\cite{Zamolodchikov:1986gt}
proved that there is an off-shell quantity $c$
that agrees with the Virasoro central charge at fixed points,
and which decreases along flows analogous to $g$ for open strings;
there have even been attempts to relate $c$ to the closed string
effective action~\cite{Tseytlin:1987bz}.
Underlying the proof of the $c$-theorem is a set of assumptions,
among which are unitarity and a gapped spectrum of $L_0$.
In the context of the nonlinear sigma model, 
this means the target space is
compact.  However, in infinite volume, the effect of the tachyon
condensate is localized to a finite region for any finite point
along the flow; since the 2d stress tensor is a continuum
normalized operator (it has an implicit factor of the volume
of space in its normalization), any localized perturbation
that might change $c$ is overwhelmed by the contribution
of spatial infinity where $c$ remains unchanged.
Of course, this is a bit too trivial; one might ask whether
what forms is a bubble of a space of lower $c$, which then grows
and takes over the whole target space in the IR.
We will see in the examples that this does not occur, but
it is an interesting open question to determine whether or not
such a scenario is possible in principle.

As mentioned in section 2.1,
there is not yet a compelling
analogue of the effective action
approach here, at least along the lines of the open
string relation $\Gamma_\ST=\ZZ_\WS^{\sst\rm disk}$.
Thus there is no direct way to relate the decay
to a decrease in the effective potential.%
\footnote{Dabholkar and Vafa%
~\cite{Dabholkar:2001wn} proposed an effective
action for orbifolds based on solutions to the $tt^*$
equations, in which the value of the effective potential is the
mass squared of the least relevant twist operator in the
chiral ring.  This proposal suffers from the fact that
it does not give the same result for $\IC^2/\IZ_{n(p)}$
and $\IC^2/\IZ_{n(n-p)}$, which as we discuss below
are isomorphic as conformal field theories.}
In particular, orbifold spaces are locally flat,
so their ADM energy always vanishes.
However, in the search for a quantity that characterizes
the flows and decreases along them,
one might ask how closed string tachyon condensation
affects the asymptotic density of closed string states
%
\be
\rho_{\cl}\sim\half g_\cl\,\Bigl(\frac{c_\eff}{3E^3}\Bigr)^{1/4}
	\exp\left[2\pi\sqrt{c_\eff E/3}\right]\ ,
\label{rhocl}
\ee
where $E=L_0+\bar L_0-\coeff1{12}c_\eff$.%
\footnote{Note that the exponential behavior is the usual Hagedorn growth
in the string spectrum, since $E_\WS=m_\ST^2$
via the Virasoro constraints.}
Typically $c_\eff=c$, the Virasoro central charge;
however, in backgrounds with throats, such as the
linear dilaton throat of an NS fivebrane source,
one can have $c_\eff<c$. 
Moreover, although as argued above $c$ is not changing
along the RG flows, it can happen that $c_\eff$
changes.  In known examples it seems to decrease%
~\cite{Kutasov:1992pf,Harvey:2001wm},
compatible with the notion of thinning of degrees of freedom.
In situations where $c_\eff({\rm IR})=c_\eff({\rm UV})=c$
such as we have for orbifolds, 
it is natural to propose that the subleading
coefficient $g_\cl$ decreases along RG flows
(and of course this is only meaningful for constant $c_\eff$).
This is the direct analogue of the open string $g$-theorem,
where the analogue of $g_\cl$ is $\Gamma_\ST^2$,
see equation \pref{rhoopen}.
Note that tachyon condensation is localized in the target
space; $\rho_\cl$ in \pref{rhocl} should count only
the {\it normalizable} closed string states.
In orbifold examples, this means we only include the
twisted sectors in the sum.  

The calculation of $\rho_\cl$ proceeds much as for the open string
case; one calculates the torus partition sum
over the Hilbert space of localized closed string states
in the limit $\tau\to0$, where the sum is dominated by
the high energy spectrum in a saddle point approximation.
On the other hand, performing a modular transformation
$\tau\to -1/\tau$ the amplitude factorizes on the lowest
energy state (the CFT vacuum) in the dual channel.
For orbifold models, one has
\bbb
\ZZ_{\rm twisted} &=& {1\over \vert \G\vert}\sum_{g\not=1,h}
        \ZZ\Bigl(\sector{h}{g}\Bigr)
\nonumber\\
\ZZ\Bigl(\sector{h}{g}\Bigr)(-1/\tau)
	&=& \ZZ\Bigl(\sector{g}{h}\Bigr)(\tau)\ .
\eee
One picks $h=1$ on the RHS of the second line
to select the contribution of the identity operator 
(\ie\ the leading contribution) as $\tau\to i\infty$.
The oscillator part is trivial; the important part is
the normalization of the zero mode in the path integral
\be
\int dp\;\delta(\RR(g)p-p)
        = {\frac{1}{ \vert \det (1-\RR(g)) \vert^2}}
\ee
where $\RR(g)$ is the rotation matrix representing 
the action of $g$ on the string coordinates,
and the integral is only over those directions
where the rotation acts nontrivially.
Plugging this result into the saddle point evaluation of 
$\rho_\cl(E)$, one finds
\be
g_\cl=\frac{1}{|\G|}\sum_{g\ne \One} 
	{\frac{1}{ \vert \det (1-\RR(g)) \vert^2}}\ .
\label{gcl}
\ee
One can show that~\cite{Harvey:2001wm}
$g_\cl$ is constant when perturbing along marginal lines
of twisted perturbations, much as in the D-brane case.

The examples that have been studied are the orbifolds
$\IC^m/\IZ_n$, acting on the complex coordinates
$(X^1,...,X^m)$ of $\IC^m$ as
\be
(X^1,\dots,X^m)\longrightarrow
	(\exp[2\pi ip_1/n]X^1,\dots,\exp[2\pi ip_m/n]X^m)\ .
\ee
Note that by choosing the generator appropriately
one can always arrange \eg\ $p_1=1$, and we will do
so in what follows.
The $\gcl$ function \pref{gcl} is then
\be
\gcl(p_1,\dots,p_m;n) =
	\frac{1}{2^{2m}\,n}\sum_{k=1}^{n-1}
	\frac{1}{\Bigl[\sin\frac{\pi p_1k}n\cdot\sin\frac{\pi p_2k}n
		\cdots\sin\frac{\pi p_mk}n\Bigr]^2}\ ,
\ee
which behaves for large $n$ and fixed $p$ as
\be
\gcl\sim\mbox{\sl const.}\,\frac{n^{2m-1}}{p_1^2p_2^2\cdots p_m^2}\ .
\ee
For example, for $\IC/\IZ_n$ one has $\gcl\sim n/12$,
and for $\IC^2/\IZ_n$ one has $\gcl\sim n^3/(720p^2)$
(where $p\equiv p_2$).
It was argued by Adams, Polchinski, and Silverstein%
~\cite{Adams:2001sv}
that the generic endpoint of tachyon condensation for
nonsupersymmetric orbifolds is flat space.
For such flows the $\gcl$ conjecture is trivially satisfied,
since the endpoint of the flow has no localized closed
string states.  However, there may be a hierarchy
of `multicritical points' of the renormalization group,
and one would like to check whether lower multicritical points
always have lower $\gcl$.  For this we must understand what theories
the RG flows lead to in the far infrared.
Extended $\NN=2$ worldsheet supersymmetry will allow us
to make exact statements about the flow structure.


\subsection{$C$/$Z_n$}

The simplest example to study is $\IC/\IZ_n$.
The precise value of $\gcl$ in this case is
\be
\gcl=\frac1n\sum_{k=1}^{n-1}\frac1{[\sin\pi k/n]^2}
	=\coeff1{12}(n-1/n)\ ;
\ee
thus, if $\gcl$ decreases under RG flow,
this should have something to do with a decrease of $n$.
In fact, one can map out the flows exactly
using the fact that while spacetime supersymmetry is broken,
one still has the $\NN=(2,2)$ worldsheet supersymmetry
generated by the worldsheet stress tensor $T$, 
supersymmetry currents $G$, $G^*$, and R-symmetry current $J$
(and their antiholomorphic counterparts).
BPS representations of worldsheet supersymmetry are {\it chiral states}
with $h=q/2$, where $h$, $q$ are the $L_0$ and $J_0$
eigenvalues.  Thus states with $q<1$ are tachyons.
Note the close parallel to the open string example of D-branes
at angles; we will return to this analogy below.
The vertex operators $\chi_q$ which create the chiral states
when acting on the CFT vacuum thus have a nonsingular operator
product expansion due to the additivity of their 
charges/scale dimensions
\be
\lim_{z'\to z}\chi_q(z)\chi_{q'}(z') = \chi_{q+q'}(z)
\ee
which determines a {\it chiral ring structure}%
~\cite{Lerche:1989uy}.
Since the operators are BPS, perturbing the worldsheet action
by $\int\! d^2z d^2\theta\, \chi +c.c$ preserves 
(the diagonal part of) $\NN=(2,2)$ supersymmetry.  

The chiral operators for $\IC/\IZ_n$ are the operators $\Sigma_{j/n}$
that create the twist ground states
\bbb
\Sigma_{j/n} &=& \sigma_{j/n}\,e^{i(j/n)(H-\bar H)}
\nonumber\\
\nonumber\\
h_j &=& \half\, \frac jn\Bigl(1-\frac jn\Bigl)
	+\half\Bigl(\frac jn\Bigr)^2 
	=\half\, \frac jn\ .
\eee
Here again $\sigma$ is the operator that implements fractional moding
on the target coordinates $X$
\be
X(e^{2\pi i}w)=e^{2\pi ij/n}\,X(w)\ ,
\ee
and the exponential of $H$
does the same for its worldsheet superpartner $\psi=e^{iH}$.
The $\NN=2$ R-current is simply $J=i\partial H$, so the
charge of these operators is $q=j/n$; they are all tachyons.
The chiral ring is
\be
\Sigma_{j/n}\Sigma_{j'/n}= 
	\cases{\Sigma_{(j+j')/n} & $j+j'<n$\cr
		& \cr
		\frac{1}{\rm Vol} e^{i(H-\bar H)} & $j+j'=n$ \cr
		& \cr
		0 & otherwise\ .}
\ee
In other words, the ring is generated by $W=\Sigma_{1/n}$,
with the relation $W^n=\frac{1}{\rm Vol}\psi\bar\psi^*\equiv\VV_\X$.
Note that we have included a factor of the volume of
the target to remind ourselves of the relative
normalizations of the localized twisted states
and the delocalized untwisted states.
One can render the volume finite by compactifying the plane
to a very large $\IP^1$ to make these statements
precise~\cite{Cecotti:1992th};
the untwisted operators decouple from the twisted
sector operators in the infinite volume limit.
We may view $W$ as a kind of winding operator around
the tip of the orbifold cone, and the chiral ring relation
as a description of the `location' of the $n-1$ twisted
vacua in the $W$-plane.

\begin{figure}[ht]
\begin{center}
\includegraphics[scale=.45]{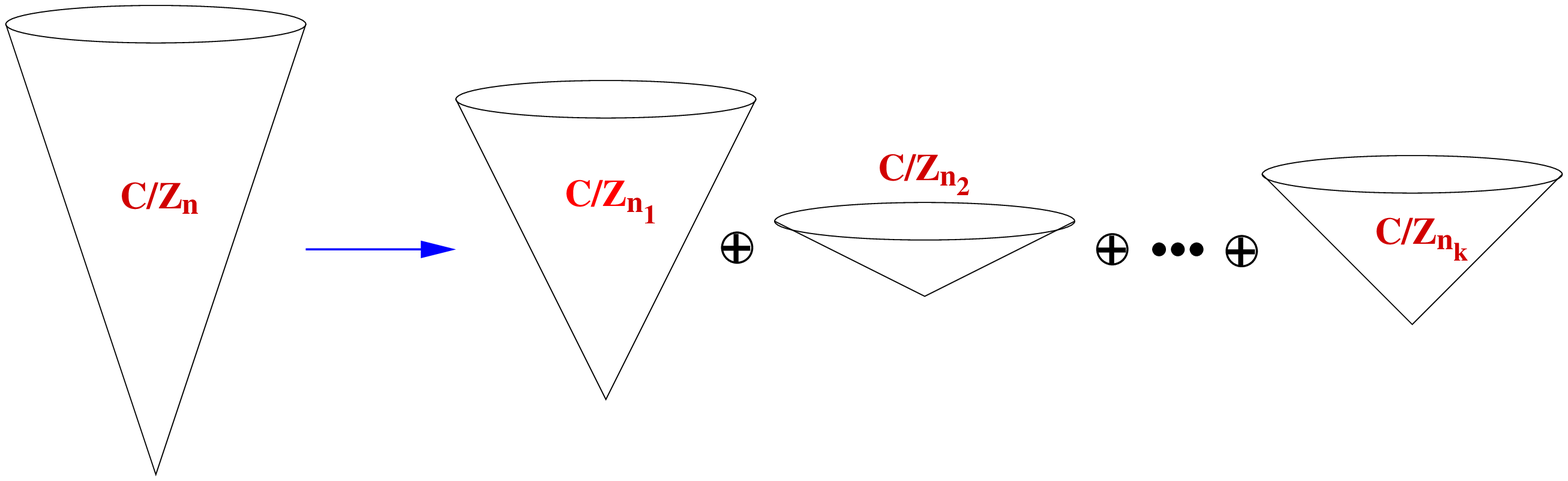}
\caption{
Tachyon condensation on $\IC/\IZ_n$ leads to
a number of disconnected spaces at the IR fixed point.
}
\label{cones}
\end{center}
\end{figure}

The next question to address is what happens when one
condenses the $\Sigma$'s:
\be
\SS_\WS=\SS_0+\Bigl(
	\lambda_j\int\! d^2zd^2\theta\,W^j+{\it c.c}\Bigr)\ .
\ee
With the $\lambda_j$ turned on, the RG flow
can deform the chiral ring structure, but cannot destroy
it since the operators are BPS and hence protected by supersymmetry.
The general deformation is
\be
W^n=\VV_\X\longrightarrow
	(W-w_1)^{n_1}\cdots(W-w_k)^{n_k} = \VV_\X\ .
\ee
with $\sum_j n_j=n$ and $w_i=w_i(\lambda)$.
The $w_i$ grow in the infrared under the renormalization group,
\ie\ the vacua decouple by becoming widely separated
in the $W$-plane.
If in the IR we focus on the neighborhood of one of the $w_i$
in the complex $W$-plane we find the chiral ring of one of
the lower orbifolds
\be
c_i(W-w_i)^{n_i}={\rm Vol.~form}\ ,
\ee
\ie\ in this part of the field space we find $\IC/\IZ_{n_i}$.
But this happens for each of the $w_i$, $i=1,...,k$,
each of which decouples from all the others in the IR.
The picture that is suggested is that of a `splitting'
of the universe into several disconnected components,
see figure \ref{cones}.
Although we have not discussed it, only one
of these vacua can be type II and therefore not
have a bulk tachyon; the rest are type 0,
due to an intriguing spontaneous breaking of
the discrete wordsheet GSO gauge symmetry under tachyon
condensation.  See~\cite{Harvey:2001wm} for details.

\begin{figure}[ht]
\begin{center}
\includegraphics[scale=.5]{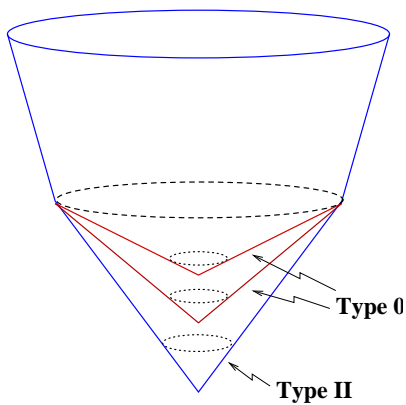}
\caption{
A possible picture of the geometry at an intermediate
stage of the decay.
}
\label{laminate}
\end{center}
\end{figure}

A possible picture of intermediate stages of the flow
is given in figure \ref{laminate}; it depicts a kind of `delamination'
of spacetime, wherein the separate cones that are forming
near the origin join onto the rest of the original cone
which is still asymptotically $\IC/\IZ_n$.%
\footnote{Note that this picture is somewhat of a cartoon;
the orbifold fixed points cannot be separating in the
spacetime parametrized by $X$, since the twist operators
respect the $U(1)$ rotational symmetry of the plane.
Rather, as the chiral ring suggests, the fixed points are separating
in the `T-dual' coordinate $W=W_1$ which is the 
field space of the twist operators.}

The quantity $\gcl$ indeed decreases along
the flows; one has
\be 
\gcl=\frac1{12}\Bigl(n-\frac1n\Bigr)\longrightarrow
	\frac1{12}\sum_{k=1}^{k} \Bigl(n_j-\frac1{n_j}\Bigr)
\ee
and indeed the IR value of $\gcl$ is smaller
than the UV value.  Thus, $\IC/\IZ_n$ decays in an interesting
way: The cone splits into several `smaller' 
(\ie\ shallower angle) cones, and the asymptotic 
density of states associated to the defect indeed
is dissipated by the flow.

\begin{figure}[ht]
\begin{center}
\includegraphics[scale=.5]{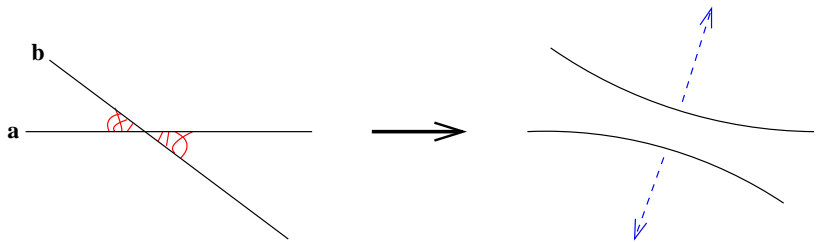}
\caption{
An open string analogue of `universe splitting'
via tachyon condensation.
This example was suggested by J. Polchinski.
}
\label{angledecay}
\end{center}
\end{figure}

Amusingly, there is actually a rather close analogue
of this splitting process in open string condensation,
namely the condensation of the tachyon on the
intersection of branes at angles of figure \ref{Dangles}.
The effect of the tachyon is to reconnect the
branes as shown in figure \ref{angledecay}.

In the IR the branes move far apart, and the fixed
point is two infinitely separated, decoupled branes.
Note also that there is an appropriate analogue
of $\gcl$ in that what is proper to count here
is the density of open string states localized
at the intersection and not in the bulk of the
brane network; these are the states that are
dissipated when the tachyon condenses.


\subsection{$C^2$/$Z_n$: The chiral ring}

The collection of $\IC^2/\IZ_n$ orbifolds provides
an even richer geometrical structure,
which is intimately connected with the resolution
of singularities in algebraic geometry.
Our main task will be to understand the structure of RG
flows, and then to check whether $\gcl$
decreases.
The group action is 
\be
(X\equiv X^1+iX^2,Y\equiv X^3+iX^4)\longrightarrow
	(\omega X,\omega^pY)\ ,
\label{znpaction}
\ee
where $p\in\IZ$, and $\omega=\exp[2\pi i/n]$.
This orbifold group is denoted $\IZ_{n(p)}$.
Note that for $p=n-1$,
the rotation is in $SU(2)$ rather than $U(2)$
so that spacetime supersymmetry is preserved;
these are the well-known $A_{n-1}$ ALE orbifolds
(\cf~\cite{Anselmi:1994sm}).

The chiral ring is built out of twist operators for 
each separate complex plane
\be
\Sigma_j=\Sigma_{j/n}^{\sst(X)}\Sigma_{\{jp/n\}}^{\sst(Y)}
\label{znptwist}
\ee
where $\{\xi\}$ denotes the fractional part of $\xi$,
$0\le\{\xi\}<1$.  

Before proceeding, a point of notation is in order.
Mathematically, a $\IZ_{n(p)}$ orbifold is defined by
the action \pref{znpaction}, and $p$ is defined
modulo~$n$, $p\sim p+n$.  However, as conformal
field theories, the orbifolds for $p$ and $-p$
are isomorphic, differing only by the field redefinition
$Y\to Y^*$.  Correspondingly, these theories contain two
sets of `BPS protected' twisted sector operators, namely 
\pref{znptwist} and 
\be
\Sigma_{j/n}^{\sst(X)}(\Sigma_{1-\{jp/n\}}^{\sst(Y)})^*
\label{caring}
\ee
which is BPS under a different linear combination
$G_\X+G^*_\Y$ of the supersymmetry currents of the component
theories.  We can call the ring of these latter operators
the $(c_\X,a_\Y)$ ring, and the ring of operators \pref{znptwist}
the $(c_\X,c_\Y)$ ring.  The operation $Y\to Y^*$
exchanges these two rings, and sends $p\to -p$.
To reconcile the mathematical $p=p_{\rm math}$, defined modulo $n$, 
and the conformal field theory $p=p_{\CFT}$,
defined modulo $2n$ but with an equivalence $p_\CFT\sim -p_\CFT$,
we will adopt the convention that $p_{\rm math}$ is
always positive, so that
\be
p_{\rm math}=n-|p_\CFT|\ ;
\ee
note that $p_{\rm math}$ and $p_\CFT$ are equivalent
modulo $n$ if we restrict ourselves to $0<p_{\rm math}<n$
and $-n<p_\CFT<0$.%
\footnote{Unfortunately, the notation of~\cite{Harvey:2001wm}
is not internally consistent in this regard;
the orbifolds labelled by $p>0$ there should actually be shifted
by $p\to p-n<0$ to agree with the above.}
We will do so in the following, and label singularities
by their value of $p_{\rm math}$.

\begin{figure}[ht]
\begin{center}
\includegraphics[scale=.45]{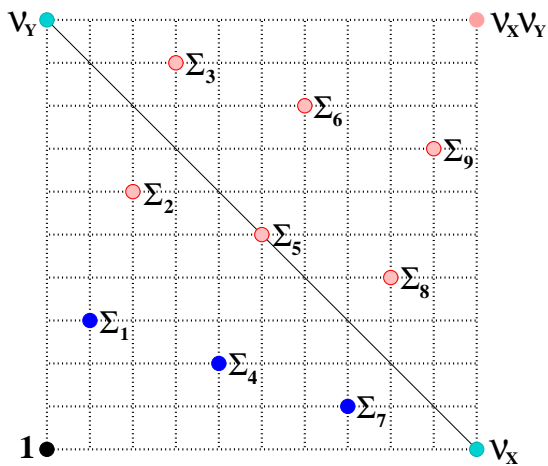}
\caption{
A plot of the R-charges of the chiral ring states
of the orbifold $\IC^2/\IZ_{n(p)}$, for $n(p)=10(3)$.
The operators $\VV_\X$ and $\VV_\Y$ have charge vectors
$(q_x,q_y)=(1,0)$ and $(0,1)$, respectively, so twisted
chiral operators have fractional charges.
The diagonal line divides the operators into
the class of irrelevant operators (above the line),
relevant operators (below the line), and marginal
operators (on the line).  The generators
of the chiral ring are denoted by solid blue dots.
}
\label{Rcharge}
\end{center}
\end{figure}

The R-charges $(q_1,q_2)$ of the chiral ring
under the $U(1)\times U(1)$ R-symmetry of the orbifold
CFT are depicted in figure \ref{Rcharge} 
for $n=10$ and $p=3$.
Note that the volume elements $\VV_{\X}$, $\VV_{\Y}$ 
of the two complex planes will always
be the maximal charge chiral operators for the
R-charges $q_x$, $q_y$ 
(since this is a property of $\IC/\IZ_n$). 
Note also that, unlike $\IC/\IZ_n$, the chiral ring
is not in general singly generated -- the charge vectors
are not all integer multiples of a single basis vector,
rather there are typically several generators.
For the example in figure \ref{Rcharge},
there are three generators:
$W_1=\Sigma_1$, $W_2=\Sigma_4$, and $W_3=\Sigma_7$.
The conformal dimension of operators is just 
$\hf(q_x+q_y)$; thus, all operators below the diagonal
line in the figure are tachyons, all operators above
the line are irrelevant, and operators on the line
are marginal.

\begin{figure}[ht]
\begin{center}
\includegraphics[scale=.6]{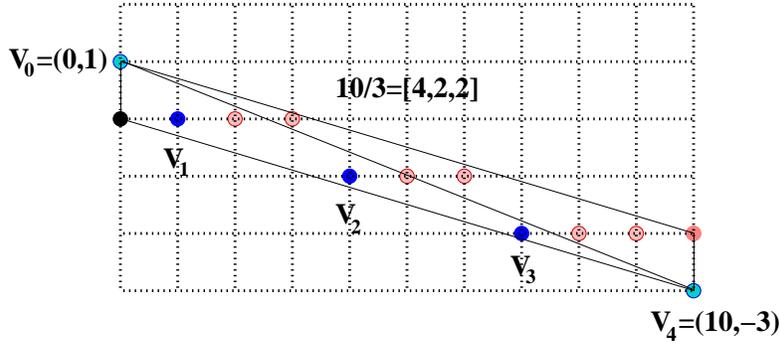}
\caption{
The chiral ring states for $n(p)=10(3)$
plotted in an integral basis
which is closely related to the resolution
of the singularity in algebraic geometry.
The vectors $\vv_i$ correspond to the ring generators.
}
\label{ring}
\end{center}
\end{figure}

It will be useful to plot the R-charges
$(q_1,q_2)$ of the $\IC\times\IC$ twist operators
in an integral basis of charges
\be
(q_x,q_y)=
	\left(j/n\;,\;j'/n=\{{pj}/{n}\}\right)
	\longrightarrow \left(j,k\equiv \coeff1n{(j'-pj)}\right)\ .
\label{chargemap}
\ee
%
This new basis is plotted in figure \ref{ring}
for $n(p)=10(3)$.  Note that the charge vectors
$(0,1)$ of $\VV_\Y$, and $(1,0)$ of $\VV_\X$,
map to $(0,1)$ and $(n,-p)$ respectively.

\begin{figure}[ht]
\begin{center}
\includegraphics[scale=.7]{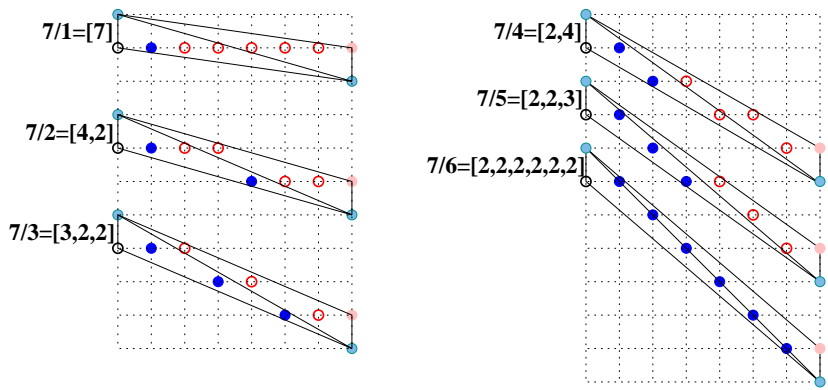}
\caption{
The chiral ring states for $n=7$, $p=1,...,6$
plotted in the integral basis.  
The black dot in the plot for each $p$ is the identity operator;
solid blue dots denote ring generators; hollow red dots
indicate the remaining ring elements.
The sequences in square brackets specify the continued fractions
of $n/p$, which will be defined later.
}
\label{pseven}
\end{center}
\end{figure}

To give a flavor of what happens as $p$ varies,
all the orbifolds for $n=7$ are plotted in figure \ref{pseven}.

As we will now show, this way of plotting the
chiral ring is directly related to the 
resolution of the $\IC^2/\IZ_{n(p)}$ singularity 
in algebraic geometry.
To see this, we need to embark on a rather lengthy digression
into algebraic geometry to prepare the necessary tools.
For a more extensive exposition, see~\cite{fulton}.


\subsection{The geometry of singularity resolution}

\begin{figure}[ht]
\begin{center}
\includegraphics[scale=.55]{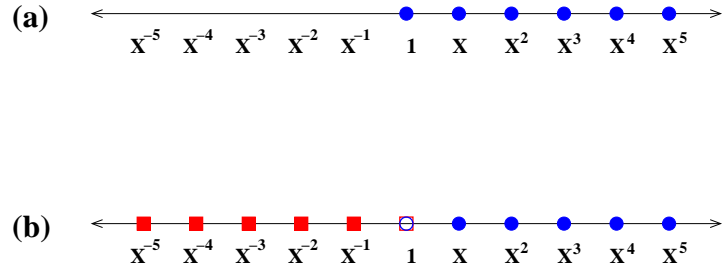}
\caption{
(a) Semi-infinite 1d lattice characterizing 
generators for the ring of holomorphic functions on $\IC$.
(b) Two such lattices patch together to describe $\IP^1$.
}
\label{onedlattice}
\end{center}
\end{figure}

Spaces in algebraic geometry are characterized by their local
rings of holomorphic functions; points are characterized
by the ideals of holomorphic functions in the ring that vanish 
at that point.
To begin, consider the complex plane $\IC$.
Its ring of holomorphic functions has a basis
$\{1,x,x^2,x^3,...\}$.  The exponents form a semi-infinite
integer lattice which is plotted in figure \ref{onedlattice}a.
A less trivial example is $\IP^1$, which must be covered by
two coordinate patches, each of which is isomorphic to $\IC$.
The corresponding functions are those holomorphic outside
the origin for one patch, and functions holomorphic away 
from infinity form the functions on the other patch.
This time, the lattice of exponents is infinite in both directions
if we superpose the sets coming from each of the two patches
(taking into account the map $X\to X^{-1}$
relating the coordinates),
see figure \ref{onedlattice}b.

\begin{figure}[ht]
\begin{center}
\includegraphics[scale=.6]{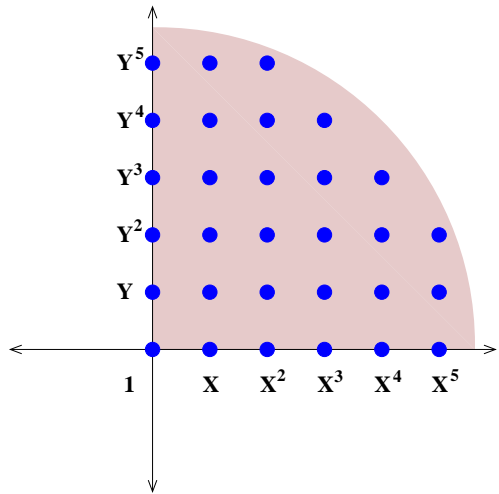}
\caption{
The lattice of generators for the
holomorphic function ring for $\IC^2$.
The shaded region is the cone $\tilde\sigma$ containing
the lattice of exponents.
}
\label{twodlattice}
\end{center}
\end{figure}

In two complex dimensions one has the obvious generalization;
the lattice for $\IC^2$ is depicted in figure \ref{twodlattice}.
Note that an isomorphic parametrization of $\IC^2$
results if we consider any $SL(2,\IZ)$ transformation
of the lattice; for instance, a ring of holomorphic
functions on $\IC^2$ can be made from polynomials
in say $X$ and $X^{-1}Y$, or from $XY^{-1}$ and $Y^{-1}$.
So in general, the ring of holomorphic functions 
on a coordinate patch is equivalent to an integer
lattice lying inside a {\it cone},
let us denote it as $\tilde\sigma$.

There is now also room for interesting structure.
In algebraic geometry, orbifolds are singular spaces;
the singularity can be {\it resolved} by excising the singular
locus and inserting a smooth space,
a procedure known as {\it blowing up}.
Rather than going directly to the singular orbifold,
let us illustrate the technology in a simpler situation
by blowing up a point on $\IC^2$, replacing it
by a $\IP^1$.  This operation is defined by the
algebraic equation
\be
X T_1 = Y T_0
\label{blowupeq}
\ee
where $(X,Y)$ are coordinates for $\IC^2$
and $(T_0,T_1)\sim (\lambda T_0,\lambda T_1)$
are homogenous coordinates for $\IP^1$.
In other words the blown up space is a surface in $\IC^2\times\IP^1$.
Locally the space looks like $\IC^2$, \ie\ $X,Y$
determine a point on $\IP^1$ through the equation \pref{blowupeq},
except at $X=Y=0$ where there is an entire $\IP^1$
since \pref{blowupeq} is automatically satisfied there.
Note that the blown up space has a different topology --
there is an additional two-cycle in the cohomology
coming from the $\IP^1$.

\begin{figure}[ht]
\begin{center}
\includegraphics[scale=.55]{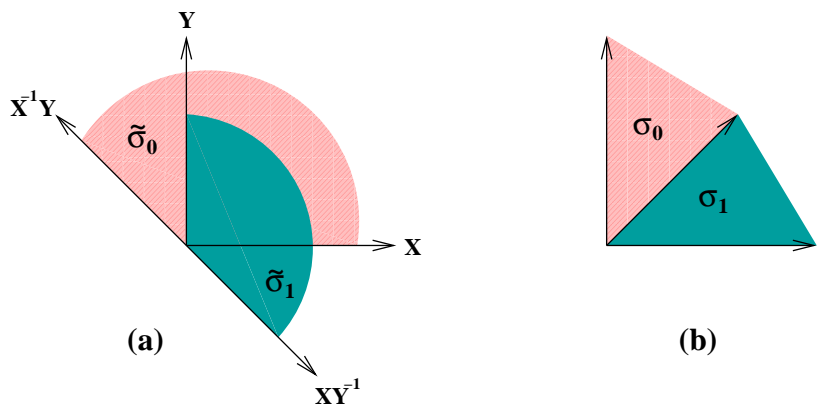}
\caption{
(a) Two cones $\tilde\sigma_0$, $\tilde\sigma_1$
contain lattices isomorphic to the $\IC^2$
lattice; these two coordinate patches combine to cover
the blowup of $\IC^2$ at a point.  The lattice 
of figure \ref{onedlattice}b for $\IP^1$,
generated as all the powers of $XY^{-1}$,
is formed by the boundaries of the two coordinate cones.
(b) The cones $\sigma_i$, dual
to the cones $\tilde\sigma_i$ containing the function
ring lattices, splice together to span the 
cone dual to the $\IC^2$ cone of figure \ref{twodlattice}.
The vector that divides $\sigma$ into $\sigma_{0,1}$
is orthogonal to the $\IP^1$ lattice contained
in $\tilde\sigma=\tilde\sigma_0\cup\tilde\sigma_1$.
}
\label{blowup}
\end{center}
\end{figure}

To cover the blowup, one needs two coordinate patches
isomorphic to $\IC^2$: A patch $U_0$, 
whose ring of holomorphic functions
is generated by $(x_0=X,y_0=X^{-1}Y=T_1/T_0)$;
and a second patch $U_1$, 
whose ring is generated by $(x_1=Y,y_1=XY^{-1}=T_0/T_1)$.
These correspond to the two patches needed to cover $\IP^1$,
as above.  On overlaps, one has the relations
$x_1=x_0y_0$ and $y_1=x_0^{-1}$.
The entire structure of coordinate patches and their
transition functions is encoded in the relationship
between the cones $\tilde\sigma_{0,1}$
containing the lattices of exponents,
see figure \ref{blowup}a.

At this point it is useful to introduce the notion of
the cones dual to the cone bounding the lattice of exponents.
It will in fact be this dual cone that relates directly
to the chiral ring when we come to consider orbifolds
and their resolution.  The cone $\sigma$ dual
to the cone $\tilde\sigma$ containing the lattice of monomial exponents 
is the set of all vectors having a positive inner product
with all the exponents in $\tilde\sigma$
\be
\vv\in\sigma\qquad{\rm IFF}\qquad 
	\vev{\vv,\uu}>0\quad \forall\ \uu\in\tilde\sigma\ .
\ee
Consider the blowup of $\IC^2$; the cones $\tilde\sigma_{0,1}$
for the two coordinate patches $U_{0,1}$
are shown in figure \ref{blowup}a, and the dual cones $\sigma_{0,1}$
are depicted in figure \ref{blowup}b.  

Note that in collection of cones $\tilde\sigma_{0,1}$
we can directly see the coordinate patches of the $\IP^1$ of the blowup
(two copies of $\IC$ parametrized by $X^{-1}Y$ and $XY^{-1}$).
Note also that the collection of dual cones $\sigma_{0,1}$
do not overlap as the $\tilde\sigma$'s do, but rather nestle
side by side; in fact, they form a {\it subdivision}
of the dual cone of $\IC^2$.  It is this feature that generalizes:
The resolution of a singularity will consist of a procedure
of successive subdivision of the dual cone $\sigma$ 
to the cone $\tilde\sigma$ of the function ring 
of the singularity, until what remains is a {\it fan}
of (dual) cones describing the coordinate patches of the
resolved space and their relationships.
We now turn to orbifold examples.


\subsection{$C^2$/$Z_n$: Geometry}

\begin{figure}[ht]
\begin{center}
\includegraphics[scale=.6]{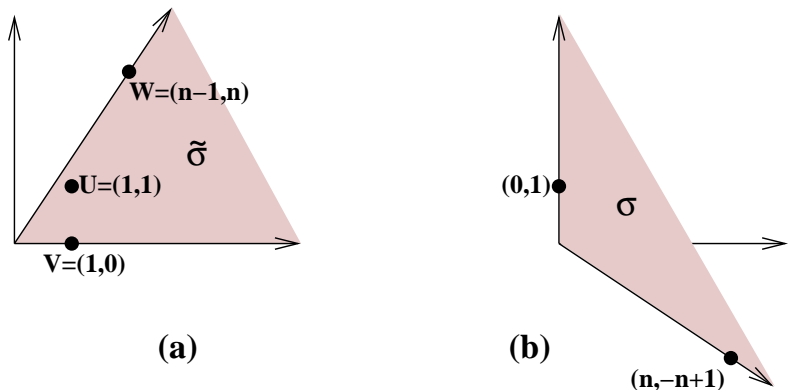}
\caption{
(a) Functions on the supersymmetric $\IZ_{n(n-1)}$ orbifold
are polynomials in $U,V,W$, related by $U^n=VW$;
the exponents of these ring generators are indicated.
The exponent of any monomial lies in the shaded cone,
and is an integer linear combination of the exponents
of $U$, $V$, and $W$.
(b) The cone $\sigma$ dual to $\tilde\sigma$.
}
\label{An}
\end{center}
\end{figure}

Consider first the supersymmetric orbifold $\IC^2/\IZ_{n(n-1)}$
which is the quotient of $\IC^2$ by the action
\be
(X,Y) \longrightarrow (\omega X,\omega^{-1}Y)
\ee
with $\omega=\exp[2\pi i/n]$ (note that $\G\subset SU(2)$
so that spacetime supersymmetry is preserved).
Holomorphic functions on $\IC^2$ invariant under this action are
\be
U=XY	\quad,\qquad
V=Y^n	\quad,\qquad
W=X^n
\ee
and in fact these generate the ring of holomorphic functions
on the orbifold.  Note however that there is a relation
\be
U^n=VW\ ;
\label{Aneq}
\ee
the surface in $\IC^3$ defined by this relation is a standard
description of this singularity in algebraic geometry.
One can make a plot of the cone of basis functions
and its dual cone, see figure \ref{An}.
The singularity of the surface is related to the fact
that the bounding vectors of the cone, 
$(1,0)$ and $(n-1,n)$,
define a parallelogram that is not of unit area
(\ie\ it encloses integral lattice points).
This means that the space is not locally like $\IC^2$
near the origin, \ie\ the function ring lattice
can't be mapped to figure \ref{twodlattice}
by an $SL(2,\IZ)$ transformation.

\begin{figure}[ht]
\begin{center}
\includegraphics[scale=.7]{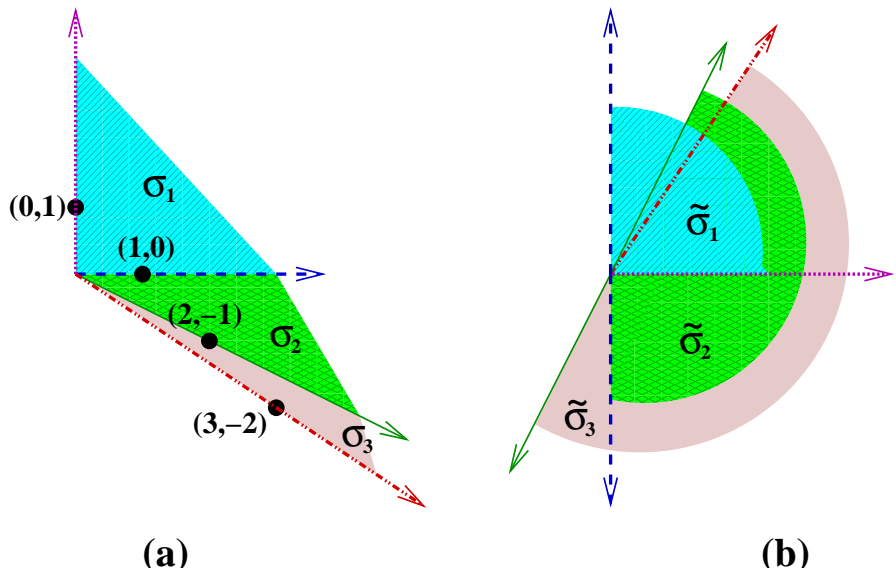}
\caption{
(a) A sequence of blowups at the singular point
subdivides the dual cone of the orbifold;
the example $n(p)=3(2)$ is depicted.
(b) The corresponding function ring cones
exhibit the $\IP^1$'s ({c.f.} figure \ref{onedlattice}b)
of the resolved space as formed by their boundaries --
the solid green line and the dashed blue line
orthogonal to the subdividing vectors $(1,0)$
and $(2,-1)$ of $\sigma$. 
}
\label{Anresolved}
\end{center}
\end{figure}

The singularity is resolved by blowing up.
The dual cone $\sigma$ is defined by basis vectors
$(0,1)$ and $(n,-n+1)$.
Choose one of the integral interior points of the
the parallelogram (the fundamental cell of the lattice)
defined by these two vectors,
and subdivide the dual cone into two sub-cones
bounded by it and either of the original two bounding vectors.
If the sub-cones are still singular (contain integral
points in the interior of their fundamental cell),
then subdivide further until the full fan of cones
correspond to nonsingular spaces.  
This subdivision is depicted in figure \ref{Anresolved}
for $n=3$.

The singularity is now resolved, since the coordinate patches
corresponding to this subdivision of the dual cone
are each isomorphic to $\IC^2$, 
and they are patched together by simple rational functions.   
For $n=3$, two successive blowups are required;
one can see the coordinate patches of the two
corresponding $\IP^1$'s being formed by the
bounding vectors of the coordinate cones in figure \ref{Anresolved},
much as in the example of blowing up $\IC^2$
considered above.  Specifically, the bounding vectors
to $\tilde\sigma_{i}$, $\tilde\sigma_{i+1}$
which are orthogonal to the dual vector $\vv_i$
that separates $\sigma_i$, $\sigma_{i+1}$
always point in opposite directions, and form the
patches of a $\IP^1$ as in figure \ref{onedlattice}b.
If we call $(x_i,y_i)$ the generators of the ring whose
corresponding cone is $\tilde\sigma_i$,
then one can read off from the figure the
(rational) coordinate maps on overlaps 
from the linear relations among the corresponding
exponent vectors.

Now we come to the relation of the resolved
singularity and the orbifold CFT.
It turns out that the set of vectors that define the 
(minimal) subdivision of the cone $\sigma$ 
which resolves the singularity
is in one-to-one correspondence with the generators
of the chiral ring of the orbifold CFT.
For $\IC^2/\IZ_{n(n-1)}$, the chiral ring consists of
the twist operators
\be
\Sigma_j=\Sigma_{j/n}^{\sst(X)}\Sigma_{1-j/n}^{\sst(Y)}\ ;
\ee
Their R-charges $(q_x,q_y)=(j/n,1-j/n)$
translate into the integer basis of equation \pref{chargemap} as
\be
(j,k) 
	= (j,1-j)
\ee
which are precisely the resolution vectors $\vv_j$!
In this example, each of the $(n-1)$ twist operators 
corresponds to a $\IP^1$ in the resolution;%
\footnote{This property does not however generalize
to other $\IZ_{n(p)}$ orbifolds, where in general
there are fewer generators of the ring
than there are twist operators.
This feature was also encountered
in $\IC/\IZ_n$, where all the twist operators
were powers of the lowest one.}
but the main point is that the data of the resolution
of the singularity is isomorphic to the data
of the orbifold chiral ring.

Thus we have the following correspondences:
\begin{itemize}
\item
The singularity is resolved by subdividing
the cone $\sigma$ (dual to the coordinate
ring of the singularity) along a set of
interior vectors $\vv_i$.  The smallest
such subdivision is the {\it minimal}
resolution of the singularity.
There is a $\IP^1$, call it $\EE_i$, blown up in 
the resolved space for each $\vv_i$.
\item
The integer coordinates $(j_i,k_i)$
of each $\vv_i$ are the R-charges in the integral basis
of a generator $W_i$ of the chiral ring of the orbifold.
The fan vectors $\vv_i$ and chiral ring generators
are in one-to-one correspondence.
The bounding vectors $(0,1)$ and $(n,-n+1)$ of the 
cone of the unresolved space
correspond to the volume elements $\VV_{\Y}$,
$\VV_{\X}$ of the two complex planes
(the only generators of the chiral ring in
the untwisted sector).
\item
The linear relations among bounding vectors
of sub-cones corresponds to relations
among the generators of the chiral ring.
For $\IC^2/\IZ_{n(n-1)}$, one has
\be
W_j^2=W_{j-1}W_{j+1}
\label{susyringrel}
\ee
which also holds for the untwisted generators
if we define $W_0\equiv\VV_{\Y}$,
$W_n\equiv\VV_{\X}$.  
\end{itemize}

We hope the reader is sufficiently convinced
of the correspondence between of the orbifold
singularity's resolution in algebraic geometry 
and the data specifying
the chiral ring of the orbifold CFT.
To actually prove the relation 
would take us too far afield; nevertheless it can
be done using the gauged linear sigma model construction of%
~\cite{Witten:1993yc,Morrison:1995fr}
(see for example~\cite{Vafa:2001ra,martmoore}).

While we have illustrated the methodology of
singularity resolution on the perhaps more
familiar example of $\IC^2/\IZ_{n(n-1)}$,
the machinery works for all the $\IZ_{n(p)}$
orbifolds.  The unresolved dual cone $\sigma$
of the singularity is bounded by the vectors
$(0,1)$ and $(n,-p)$;%
\footnote{Related to this is the fact that
the invariant functions can, as in the case of
figure \An a, be plotted so as to lie in the cone
$\tilde\sigma$ bounded by $(1,0)$ and $(p,n)$.}
note that these are the
same as the charges $(j,k)$ of the chiral
ring elements $\VV_\Y$ and $\VV_\X$ of the orbifold,
see equation \pref{chargemap}.
The resolution of the singularity is accomplished
by subdividing the cone $\sigma$ until all of the
sub-cones have unit area fundamental cell;
each subdividing vector of the minimal such resolution
corresponds to a generator of the chiral ring
of the orbifold, and the components of the vector
determine the R-charges of the ring operator
via \pref{chargemap}.  Any three successive vectors
$\vv_{i-1}$, $\vv_i$, $\vv_{i+1}$
in the fan obey a linear relation 
\be
-a_i\,\vv_i+\vv_{i-1}+\vv_{i+1}=0\quad,\qquad i=1,...,r \ ,
\label{linrel}
\ee
with integer $a_i$,
corresponding to a relation in the chiral ring
\be
W_i^{a_i}=W_{i-1}W_{i+1}
\ee
generalizing \pref{susyringrel},
again defining $W_{0,r+1}$ as $\VV_{\Y,\X}$, respectively.

\begin{figure}[ht]
\begin{center}
\includegraphics[scale=.4]{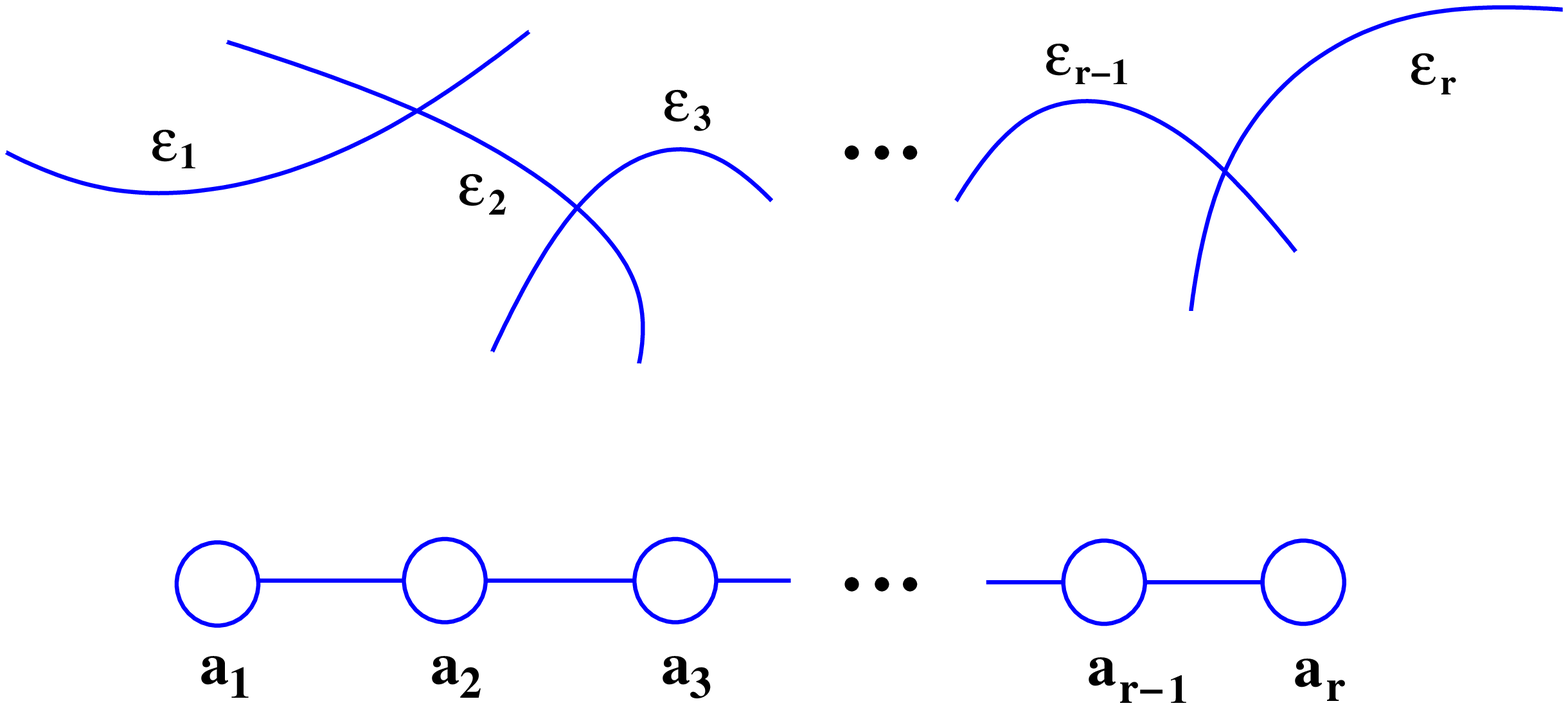}
\caption{
The intersection structure of the $\IP^1$'s $\EE_i$ of the
resolved space (schematically depicted in the upper figure)
is conveniently encoded in a Dynkin-like diagram (the lower figure).
The diagonal entries of the intersection matrix --
the self-intersection numbers of the $\EE_i$ --
label the nodes of the diagram.
}
\label{resolution}
\end{center}
\end{figure}

The coefficients in \pref{linrel} code the
intersection matrix of the $\IP^1$'s of
the resolution; namely, the $\IP^1$ corresponding
to $\vv_i$ intersects the ones corresponding to the
adjacent fan vectors $\vv_{i-1}$ and $\vv_{i+1}$
each once, and has self-intersection $-a_i$.
Note that these intersection numbers form
the Cartan matrix of $A_{n-1}$ in the case of
$\IC^2/\IZ_{n(n-1)}$, which is one reason why
the supersymmetric orbifolds are called $A_n$
singularities.  It is convenient to use the
notation of Dynkin diagrams even in the more
general case of $\IZ_{n(p)}$, since the
structure of their resolutions
only differs in the self-intersection numbers
of the $\IP^1$'s; see figure \ref{resolution}.

An amusing and somewhat magical fact is that 
the integers $a_i$ are encapsulated by the 
{\it continued fraction expansion} of $n/p$
\be
  \frac np=~a_1-\frac{1}{~a_2-\frac{1}{~a_3-\frac{1}{~\ldots~-1/a_r}}}
        ~\equiv~ [a_1,a_2,a_3,\ldots,a_r]\qquad.
\ee
For example, in the supersymmetric orbifold
$p=n-1$ one has
\be
\frac{n}{n-1} = [2,2,...,2]
\qquad\quad({\eg}\quad\frac43=2-\frac{1}{2-1/2}=[2,2,2])\ .
\ee
In fact, there is more magic: The `dual' continued fraction 
\be
\frac{n}{n-p} = [b_2,...,b_{\ell-1}]
\ee
codes the relations among the monomials 
$\wp_1,...,\wp_\ell$ generating the
ring of invariant functions on the quotient $\IC^2/\IZ_{n(p)}$ as 
\be
\wp_i^{b_i}=\wp_{i-1}\,\wp_{i+1}\quad,\qquad i=2,...,\ell-1\ .
\label{embedding}
\ee
These $\ell-2$ equations embed the singularity
as a surface in $\IC^\ell$, generalizing equation \pref{Aneq}.
Assigning vectors $\uu_1=(1,0)$ to $\wp_1$
and $\uu_2=(1,1)$ to $\wp_2$ as in figure \An a,
leads to an increasing set of vectors culminating in
$\uu_\ell=(p,n)$.%
\footnote{This proves the claim in the previous footnote.}
The $\G$-invariant monomials are given explicitly as follows:
Let 
\bbb 
\ttt_1 &=& (t_1,t'_1)=(n,0)
\nonumber\\
\ttt_2 &=& (t_2,t'_2)=(n-p,1)
\label{invtexpts}\\
\ttt_{i+1} &=& b_i\ttt_i-\ttt_{i-1}
\quad,\qquad i=3,...,\ell\ .
\nonumber
\eee
Then the ring of invariants is generated by
$\wp_i=X^{t_i}Y^{t'_i}$.

There is an interesting `duality' or `mirror symmetry'
of the $\IZ_{n(p)}$ and $\IZ_{n(n-p)}$ orbifolds.  
As was noted in section 3.2,
in conformal field theory $p$ and $-p$ 
are related by the exchange $Y\leftrightarrow Y^*$
which induces an exchange of the $(c_\X,c_\Y)$
and $(c_\X,a_\Y)$ rings.  One can then check that
the $(c_\X,c_\Y)$ ring of the $\IZ_{n(p)}$ orbifold
is the $(c_\X,a_\Y)$ ring of the $\IZ_{n(n-p)}$
orbifold, and vice versa.  
A given orbifold singularity can be
smoothed in two different ways -- by {\it resolving}
it through a K\"ahler deformation as above,
or by {\it deforming} the equations \pref{embedding}
which define it as a hypersurface in $\IC^\ell$.
These two types of smoothings are
related by exchanging $p$ and $n-p$,
as for instance the data that defines them are
given by the dually related continued fractions
for $n/p$ and $n/(n-p)$.
A given $\IZ_{n(p)}$ orbifold
has both sets of deformations in it 
(at least before GSO projection, \ie\ in type 0).  
The GSO projection
\be
H_1\to H_1+p\pi\quad,\qquad H_2\to H_2-\pi
\ee
keeps some of each ring, namely
the $(c_\X,c_\Y)$ states with $[jp/n]\in 2\IZ+1$
and the $(c_\X,a_\Y)$ states with $[jp/n]\in 2\IZ$
(here $[\xi]=\xi-\{\xi\}$ denotes the integer part of $\xi$).
It is only for the supersymmetric orbifold that the 
entire $(c_\X,a_\Y)$ ring is projected out and
the entire $(c_\X,c_\Y)$ ring is preserved by the GSO projection.
Note that this mirror symmetry is different from the 
usual one (\cf~\cite{Greene:1991iv}), 
which is T-duality of the $\NN=(2,2)$
$U(1)$ R-current $J$; that operation exchanges chiral and 
twisted chiral fields (\ie\ switching chiral and antichiral for
left-movers, keeping right-movers fixed) for {\it all} the superfields.



\subsection{$C^2/Z_n$: RG Flows}

In string theory, one does not
actually have to blow up the geometrical size
of the $\IP^1$'s in order to resolve the singularity,
as one must in algebraic geometry.  
The orbifold is a completely nonsingular CFT.
The size $V_i$
of the $\ith$ cycle comes complexified as
$V_i+i B_i$ where $B_i$ is the NS B-flux through the cycle.  
Aspinwall showed~\cite{Aspinwall:1994ev,Aspinwall:1994xz}
that the supersymmetric orbifolds have $1/n$ unit of $B$-flux turned on
through each $\IP^1$, thus blowing up the curves 
in a non-geometrical `imaginary' direction
even though the cycles are collapsed to zero size.
This $B$-field resolves the singularity in a way
that strings detect (for instance, it gives finite
action to worldsheet instantons wrapping the $\IP^1$).
One imagines that a similar story transpires for the other
$\IZ_{n(p)}$ orbifolds -- that even though the $\IP^1$'s
that resolve the singularity are collapsed to zero size,
a nonzero $B$-flux through them ensures the regularity
of string propagation on the orbifold.

In the perturbed worldsheet action
\be
\SS_\WS=\SS_0+\Bigl(\lambda_j\int\! d^2zd^2\theta\,W_j+{\it c.c.}\Bigr)\ ,
\label{pertact}
\ee
the complex parameter $\lambda_j$ is some function
of the complexified K\"ahler parameter of the corresponding $\IP^1$.
For the supersymmetric case $\IZ_{n(n-1)}$, 
all twist operators correspond
to different curves, and they are all marginal operators.
The resulting family of CFT's are the Eguchi-Hanson ALE spaces
(\cf~\cite{Anselmi:1994sm}).
For general $\IZ_{n(p)}$, not all the chiral ring elements
correspond to different curves in the resolution;
for $p\ne n-1$, one has $r<n-1$.  Moreover, the ring elements corresponding
to the curves of the algebraic resolution are relevant operators;
perturbing the action \pref{pertact}
still blows up the curves geometrically,
but now the couplings $\lambda_j$ grow along RG flows
and thus the size of the corresponding 
curve increases as we flow to the IR.

\begin{figure}[ht]
\begin{center}
\includegraphics[scale=.45]{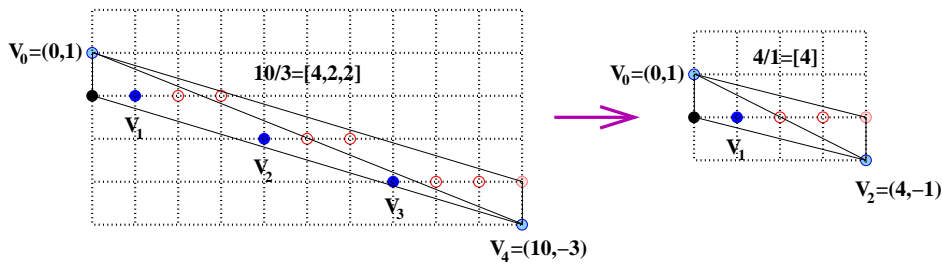}
\caption{
The RG flow that blows up to infinite size
the middle $\IP^1$ of the resolved $10(3)$ singularity,
yields a daughter singularity for which the 
fan vector corresponding to the $\IP^1$
becomes the bounding vector of the dual cone
of the daughter.
}
\label{orbdecay}
\end{center}
\end{figure}

We are now finally prepared to discuss the RG flows
for $\IC^2/\IZ_{n(p)}$.
Blowing up the $\ith$ $\IP^1$ of the minimal resolution
to infinite volume turns that $\IP^1$ into $\IC$.
Thus we expect that the corresponding operator $W_i$
becomes the volume form $\VV$ for a noncompact
direction of a new singularity, with the corresponding
charge vector $(j_i,k_i)$ being the bounding vector
of its dual cone, see figure \ref{orbdecay}.

\begin{figure}[ht]
\begin{center}
\includegraphics[scale=.55]{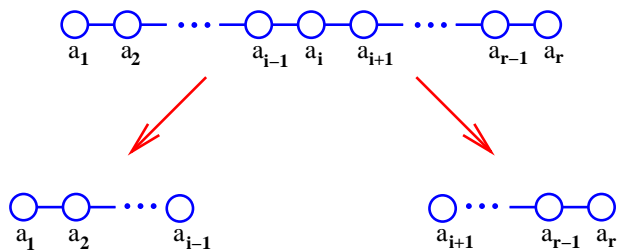}
\caption{
The Dynkin diagram of the singularity
splits by the deletion of the node corresponding
to the curve being blown up to infinite size.
As in figure \ref{cones}, spacetime splits into disconnected components.
}
\label{split}
\end{center}
\end{figure}

The other half of the split fan of cones can be converted into the
canonical form with a bounding vector at $(0,1)$
by an $SL(2,\IZ)$ transformation of the lattice.
For the example $n(p)=10(3)$ shown in the figure,
the other singularity is $n(p)=2(1)$.
Note that this is precisely what we would get by splitting 
apart the Dynkin diagram (figure \ref{resolution})
of the singularity by deleting the node corresponding to $\vv_2$, 
see figure \ref{split}.
This makes sense since this
curve is leaving the set of finite volume elements
of the middle homology.
The limit $\lambda_i\to\infty$ effectively imposes a
Lagrange multiplier constraint $W_i=0$,
as can be seen from the action \pref{pertact}
-- just as the strict infinite volume limit
removes $\VV_{\X,\Y}$ from the normalizable
elements of the chiral ring.

More generally, one could perturb the action by
a chiral ring element that does not correspond to one
of the fan vectors $\vv_i$ of the minimal resolution.
What happens then?  
In geometry, one can by a succession of further blowups
make that operator correspond to a $\IP^1$ of a
non-minimal resolution of the singularity.
The additional blowups correspond to adding more
vectors to the fan, further subdividing
the already nonsingular dual cones $\sigma_i$
(very much like blowing up $\IC^2$ as we did above).
The effect of a single blowing up operation is
to modify the data of the continued fraction expansion,
which defines the relations among the fan vectors
via \pref{linrel}, as follows:
\be
[a_1,\cdots, a_k]\to
  [a_1,\ldots,(a_{i-1}+1)\,,1,\,(a_{i}+1),\ldots,a_r]\ .
\label{blowupfrac}
\ee
One can check that this continued fraction sequence
defines the same fraction $n/p$, and indeed
has the effect of subdividing the $\ith$ cone $\sigma_i$
while leaving the others intact.
The effect of flowing to the IR under the perturbation
corresponding to this non-minimal curve should be to split 
this expanded Dynkin diagram in two by
deleting the appropriate node.  Similarly, any operator
whose curve can be put into the Dynkin diagram by
a sequence of blowups ought to work the same way.
Any chiral operator that is not a power of another
can be treated in this manner.

\begin{figure}[ht]
\begin{center}
\includegraphics[scale=.9]{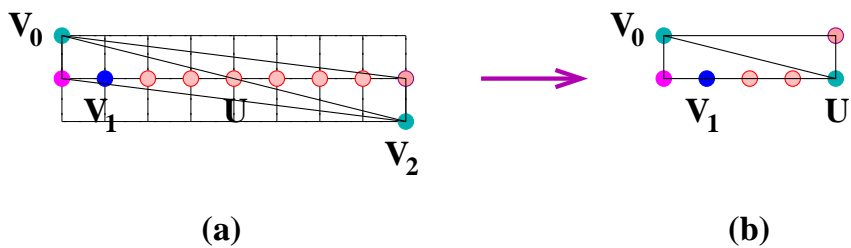}
\caption{
(a) Chiral ring/toric fan for $n(p)=8(1)$.
(b) Chiral ring/toric fan for $\IC/\IZ_4\times \IC$,
obtained by blowing up the curve corresponding to $\uu=4\vv_1$, 
associated to the chiral operator $W^4$ 
(the chiral operator $W$ having R-charge vector $\vv_1$
generates the chiral ring).
}
\label{resfantwo}
\end{center}
\end{figure}

Finally, what if the chiral operator we perturb by is not a generator
of the minimal resolution, or any other resolution that
can be obtained from it by a sequence of blowups?
Consider for simplicity the case of $n(p)=2\ell(1)$, $\ell\in\IZ$.
Call the generator of the chiral ring $W$; its charge vector
is labelled $\vv_1$ in figure \ref{resfantwo}a.
Here we understand what the $W^\ell$ perturbation does,
since the orbifold can be thought of as the
supersymmetric orbifold $\IC^2/\IZ_{2(1)}$, further
orbifolded by $\IZ_{\ell}$; the marginal operator
is the one that blows up the $\IC^2/\IZ_{2(1)}$
into an $A_1$ ALE space, whose geometry is that of
the cotangent space of the sphere $T^*\IP^1$.
Parametrize $T^*\IP^1$ by coordinates $x_\pm$, $p_\pm$ related to
the standard coordinates $Z_1$, $Z_2$ of $\IC^2$ by
\bbb
  x_+ = Z_1/Z_2 \qquad & & \qquad p_+ = Z_2^2\nonumber\\
  x_- = Z_2/Z_1 \qquad & & \qquad p_- = - Z_1^2\ .
\label{coordrel}
\eee
Then $\IZ_{2\ell(1)}$ does not act on $x_\pm$, and acts as
$\IZ_\ell$ on $p_\pm$, so the infinitely
blown up theory is $\IC/\IZ_\ell\times\IC$.
%
But we get exactly this picture in the resolution diagram,
see figure \ref{resfantwo}.
We recognize along the horizontal axis the toric cone
of $\IC/\IZ_\ell$; the cone is singular because the
cone of $\IC$ is not generated by a primitive vector.
Rather, the invariant functions on the space are functions
of $Z_1^\ell$ and $Z_2$.  The picture clearly generalizes
to any perturbation not `inherited' from the
ALE space, \ie\ any of the relevant
operators $W^k$, $k<\ell$, in the chiral ring of $\IC^2/\IZ_{2\ell(1)}$;
we expect that in the IR of the RG flow
we arrive at the target space $\IC/\IZ_k\times \IC$.

Thus, a general picture arises for
the perturbation of an arbitrary resolution
fan by an arbitrary vector $\vv_j$ corresponding to a
relevant perturbation $\Sigma_j$ in the chiral ring; 
one splits the cone $\sigma$ into
two cones $\sigma_1$, $\sigma_2$
along the ray generated by the vector $\vv_j$,
and spacetime splits in the infrared of the RG flow
into the orbifolds whose singularities are specified by
$\sigma_1$ and $\sigma_2$.



\subsection{The $g_{cl}$ conjecture}

What about $\gcl$?  As mentioned above,
at large $n$ and fixed $p$ one has
$\gcl\sim n^3/(720 p^2)$.  One needs to check whether
the conjectural picture of RG flows, obtained from
deleting a node on the Dynkin diagram (figure \ref{resolution})
of the singularity, leads to daughter singularities
whose $n'(p')$ satisfies the conjectured
decrease in the asymptotic density of localized
closed string states.
Assuming the picture of the flows suggested by geometry
is valid, it is straighforward to check that the
generic flow leads to a decrease in $\gcl$.
The integer coordinates $(j,k)$ of the perturbing vector
become $(n',p')$ of one of the daughter singularities,
and in any given example one can find the values of $(n'',p'')$
of the other daughter (this is a question of performing
the $SL(2,\IZ)$ transformation that puts the other
sub-cone into the standard position with one of
the bounding vectors at $(0,1)$; then the other bounding
vector is at $(n'',-p'')$).  Unless the splitting vector
is very near the boundaries of the original cone $\sigma$,
the daughters will have a much smaller $n'$ and $n''$,
and hence $\gcl$ will decrease.

\begin{figure}[ht]
\begin{center}
\includegraphics[scale=.53]{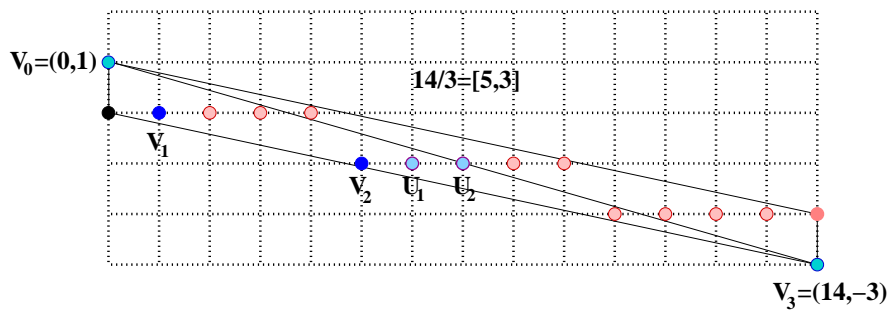}
\caption{
The flow
$2\ell(3)\longrightarrow \ell(\ell-3)\oplus \ell(1)$
involves a sequence of blowups of the sort described
by equation \pref{blowupfrac}, in order to put
the curve corresponding to the perturbing operator
into the Dynkin diagram of the resolution.
Here the vectors $\vv_i$ give the subdivision of the cone
corresponding to the minimal resolution of the singularity;
the $\uu_j$ specify the further subdivision that
adds the extra curves to the Dynkin diagram 
(in this example, $\uu_2$ corresponds
to the perturbing marginal operator).
}
\label{twolthree}
\end{center}
\end{figure}

There is however a class of candidate counterexamples
(first discussed in~\cite{Adams:2001sv}), where the splitting vector is
near the edge of the cone $\sigma$, namely
$2\ell(3)\longrightarrow \ell(\ell-3)\oplus \ell(1)$.%
\footnote{In fact, one can check using the above rules for
splitting fans via infinite blowups, that any flow
$n(p)\to n'(1) \oplus n''(n''-p)$,
with $n'=\coeff{2n}{p+1}$ and $n''=\coeff{p-1}{p+1}n$,
violates the $\gcl$ conjecture
(here we are assuming that $p+1$ divides $n$).
All these flows are however subject to
the same caveats discussed below.}
Here one has
\bbb
\gcl({\rm UV})&\sim&\frac 1{720}\frac{(2\ell)^3}{9}
\nonumber\\
\gcl({\rm IR})&\sim&\frac 1{720}\Bigl(\frac{\ell^3}9+\frac{\ell^3}1\Bigr)\ ,
\eee
\ie\ larger by $2/9$ in the IR relative to the UV.
This flow is marginal, and so $\gcl$ is constant
along the flow, but if this flow works as advertised, one
can construct flows by only slightly relevant
operators (nearby to the marginal one on the
Dynkin diagram) which also violate the conjecture.
There is some evidence, however, that this particular
conjectured flow is very peculiar, and that perhaps
we don't understand the correspondence between
geometry and worldsheet field theory as well as
we'd like to think.  For instance, to put the
perturbing operator into a Dynkin diagram of
a (nonminimal) resolution, one must blow up 
many (of order $\ell/3$) additional $\IP^1$'s
beyond those needed for the minimal resolution,
see figure \ref{twolthree}.

In other words, the continued fraction of
the UV singularity is $[m+1,3]$ for, say, $2\ell=3m+2$
(the case $2\ell=3m+1$ is similar),
while the two daughter singularities have continued fractions
$[2,...,2,4]$ and $[\ell]$, where the number of curves with $a_i=2$
is $(\ell+2)/3$.
All these curves are associated to relevant perturbations,
and it is not immediately obvious that they are not
being blown up along the flow as well, leading to
a different picture of the IR fixed point that would
not contradict the $\gcl$ conjecture.
An argument against this point of view is that,
along the flow by the marginal perturbation,
the geometry is that of a $\IZ_\ell$ orbifold
of $T^*\IP^1$, much as at the end of the previous subsection.
Thus it might be that the extra curves are secretly being
blown up by the marginal perturbation so that the
resulting conformal field theory is nonsingular.

If these orbifolds prove to be valid counterexamples
to the $\gcl$ conjecture, it is actually somewhat more
interesting than if they do not.  Where are the
additional localized closed string states coming from?
Naively one expects that in the description of the marginal
flow as a $\IZ_\ell$ orbifold of $T^*\IP^1$,
there are states localized at the $\IZ_\ell$ fixed points
at the north and south poles of the $\IP^1$,
as well as normalizable states associated to the $\IP^1$;
the latter should disappear from the normalizable
spectrum at infinite volume, leaving only the localized
orbifold states, and it does not seem as though we
have gained any states in the process.


\begin{acknowledgments}
Thanks to
J. Harvey,
D. Kutasov,
and
G. Moore
for enjoyable and stimulating collaborations
that led to the results presented here;
and also to
B. Craps,
J. Polchinski,
and
E. Silverstein
for helpful discussions and correspondence.
Thanks to the organizers of the school for all their efforts.
This work was supported by DOE grant DE-FG02-90ER-40560.
\end{acknowledgments}


\end{document}